\documentclass[5p]{elsarticle}
\usepackage{amsmath, amssymb, graphicx}
\usepackage[utf8]{inputenc}
\usepackage{lipsum}
\usepackage{overpic}
\usepackage{wasysym}
\usepackage[dvipsnames]{xcolor}
\usepackage{subcaption}
\usepackage{siunitx}
\usepackage{breakurl}
\usepackage[version=4]{mhchem}
\usepackage{listings}
\usepackage{minted}
\usepackage{supertabular,booktabs}
\usepackage[colorlinks = true,linkcolor = blue,urlcolor  = blue, citecolor = blue]{hyperref}
\usepackage[acronym,nonumberlist,nogroupskip]{glossaries-extra}
\usepackage{gensymb}
\usepackage{multirow}
\usepackage{stfloats}

\def\cm{c_\mathrm{m}}

\newcolumntype{L}[1]{>{\raggedright\arraybackslash}p{#1}}
\newcolumntype{C}[1]{>{\centering\arraybackslash}p{#1}}
\newcolumntype{R}[1]{>{\raggedleft\arraybackslash}p{#1}}

\begin{document}

\begin{frontmatter}
\title{Multiphysics tritium transport modelling of the ARC breeding blanket with FESTIM}

\cortext[mycorrespondingauthor]{Corresponding author}
\ead{darkj385@mit.edu}

\author[MIT]{James Dark\corref{mycorrespondingauthor}}
\author[ORNL]{Arpan Sircar}
\author[ORNL]{Jin Whan Bae}
\author[Simula]{J\o rgen S. Dokken}
\author[MIT]{Rémi Delaporte-Mathurin}
\address[MIT]{Plasma Science and Fusion Center, Massachusetts Institute of Technology, Cambridge, MA 02139, USA}
\address[ORNL]{Oak Ridge National Laboratory, Nuclear Energy and Fuel Cycle Division, 1 Bethel Valley Road, Oak Ridge, TN 37830, USA}
\address[Simula]{Dept. of Numerical Analysis and Scientific Computing, Simula Research Laboratory, Oslo, Norway}

\begin{abstract}
    Accurate prediction of tritium behaviour in molten salt breeding blankets is essential for the design and safe operation of ARC-class fusion reactors.
    This work presents a fully open-source, component-scale multiphysics framework for modelling tritium transport in an ARC liquid immersion blanket. 
    Neutron transport, thermal hydraulics, and hydrogen isotope transport are coupled using OpenMC, OpenFOAM, and FESTIM, leveraging dedicated tools enabling direct transfer of spatially resolved fields between solvers. 
    Assuming a zero inlet concentration, steady-state simulations predict a total tritium inventory of approximately \SI{243}{mg}, with the blanket reaching steady-state tritium throughput within approximately \SI{30}{min}, which is of a similar order to previous system-level estimates.
    The results show that tritium transport is dominated by turbulence-enhanced diffusion, with strong localisation in flow stagnation regions and reduced accumulation in highly turbulent zones. 
    Sensitivity analyses indicate that predicted inventories are governed primarily by the numerical stabilisation scheme, with only a modest dependence on the turbulent Schmidt number.
    The proposed workflow provides a transparent and extensible basis for high-fidelity analysis of tritium transport in ARC-class breeding blankets.
\end{abstract}

\begin{keyword}
ARC, breeding blanket, tritium, multiphysics, FESTIM
\end{keyword}

\end{frontmatter}

\section{Introduction}
An integral component of the ARC (affordable, robust, compact) fusion reactor concept is its breeding blanket, which is responsible for tritium production and recovery to sustain the fuel cycle. 
In the ARC concept, this function is fulfilled by a liquid immersion blanket of FLiBe salt, which simultaneously serves as a breeder, coolant, and neutron moderator~\cite{sorbom_arc_2015}. 
FLiBe consists of two moles of LiF and one mole of BeF$_2$.
The use of a flowing molten salt introduces strong coupling between neutronics, thermal hydraulics, and hydrogen isotope transport, making predictive modelling of tritium behaviour a key challenge during the early stages of reactor design. 
At this stage, modelling efforts are not primarily intended to resolve all physical phenomena in detail but rather to support rapid design iteration by identifying dominant transport pathways, inventories, and sensitivities across candidate configurations.
To our knowledge, there are no published targets or constraints for tritium metrics (e.g. concentrations, inventories, throughput). 

Tritium control is a central consideration in the design of breeding blankets because of the high mobility of hydrogen isotopes and their propensity to migrate across material interfaces.
During reactor operation, tritium generated within the blanket can be transported by diffusion and advection, permeate into structural components, and accumulate in regions beyond the intended extraction systems~\cite{dark_influence_2021}. 
These processes directly influence the distribution of tritium inventory and extraction efficiency and must therefore be considered when assessing candidate blanket layouts and operating conditions.
Predictive modelling of tritium transport pathways and inventories is thus an essential tool for informing early design choices and identifying configurations that minimise retention and losses due to permeation.

To date, detailed studies of tritium behaviour in ARC-like systems at the component scale remain limited~\cite{ferrero_preliminary_2022}. 
Experimental investigations have addressed key aspects of tritium chemistry in FLiBe~\cite{simpson_quantitative_2006, hara_interactions_2006, fukada_reaction_2007}, interactions between molten salts and structural materials~\cite{keiser_compatibility_1979, terai_compatibility_2001}, and tritium generation and accountancy in fluoride salt systems~\cite{dolan_experimental_2023, ferry_libra_2023, delaporte-mathurin_advancing_2025}.
However, integrated simulations that represent the coupled multiphysics environment of breeding blanket components---combining neutron transport, heat transfer, and hydrogen isotope diffusion---remain scarce.
Existing modelling efforts have primarily focused on alternative blanket configurations rather than liquid immersion concepts~\cite{candido_integrated_2020, candido_novel_2021, dark_influence_2021, alberghi_magneto-convective_2020, zhou_design_2017}, reducing their relevance to ARC-class designs.

Preliminary modelling of tritium transport in ARC-relevant molten salt blankets has been demonstrated by Ferrero et al.~\cite{ferrero_preliminary_2022}, who coupled computational fluid dynamics (CFD) with hydrogen isotope transport to estimate steady-state tritium inventories, albeit using an axisymmetric 2D representation of the blanket.
In parallel, the Fusion Energy Reactor Models Integrator (FERMI) project at Oak Ridge National Laboratory developed a modular multiphysics workflow for fusion reactor components that integrates neutronics, thermal, and fluid simulations~\cite{badalassi_fermi_2023}. 
The FERMI workflow, as part of the Fusion REactor Design and Assessment (FREDA) project~\cite{freda_2025}, began to address tritium transport within the fluid using a passive scalar transport approach but did not couple this transport into the surrounding solid structures.
Together, these efforts highlight both the feasibility of tritium modelling in ARC-like systems and the need for a unified workflow that incorporates hydrogen isotope transport within an integrated, design-oriented framework.

To address these challenges, the present work introduces a modular modelling workflow for tritium transport in an ARC-relevant molten salt breeding blanket.
This workflow is built around FESTIM~\cite{dark_festim_2026}, a dedicated solver for hydrogen isotope transport. 
The objective of this study is to demonstrate a flexible framework suitable for early-stage design studies rather than to provide a fully resolved representation of a final blanket geometry. 
Multiphysics input data are obtained from established external tools within the FERMI framework, with OpenMC~\cite{romano_openmc_2015} used for neutron transport and OpenFOAM~\cite{noauthor_openfoam_nodate, greenshields_notes_2022} used for thermo-fluid dynamics. 
To support the integration of these simulations, two dedicated packages have been developed: \texttt{foam2dolfinx}~\cite{dark_festim-devfoam2dolfinx_2025}, which converts OpenFOAM results into formats compatible with DOLFINx~\cite{baratta_dolfinx_2023} and FESTIM, and \texttt{openmc2dolfinx}~\cite{dark_festim-devopenmc2dolfinx_2025}, which enables the import of OpenMC outputs. 
Together, these developments establish a transparent and extensible workflow for assessing trends in tritium production, transport, and inventory in simplified ARC-relevant geometries as design assumptions evolve.

\section{Methodology}
\begin{figure}[H]
    \centering
    % trim={<left> <lower> <right> <upper>}
    \includegraphics[width=\linewidth, trim={5.5cm 3cm 7.0cm 4.8cm}, clip]{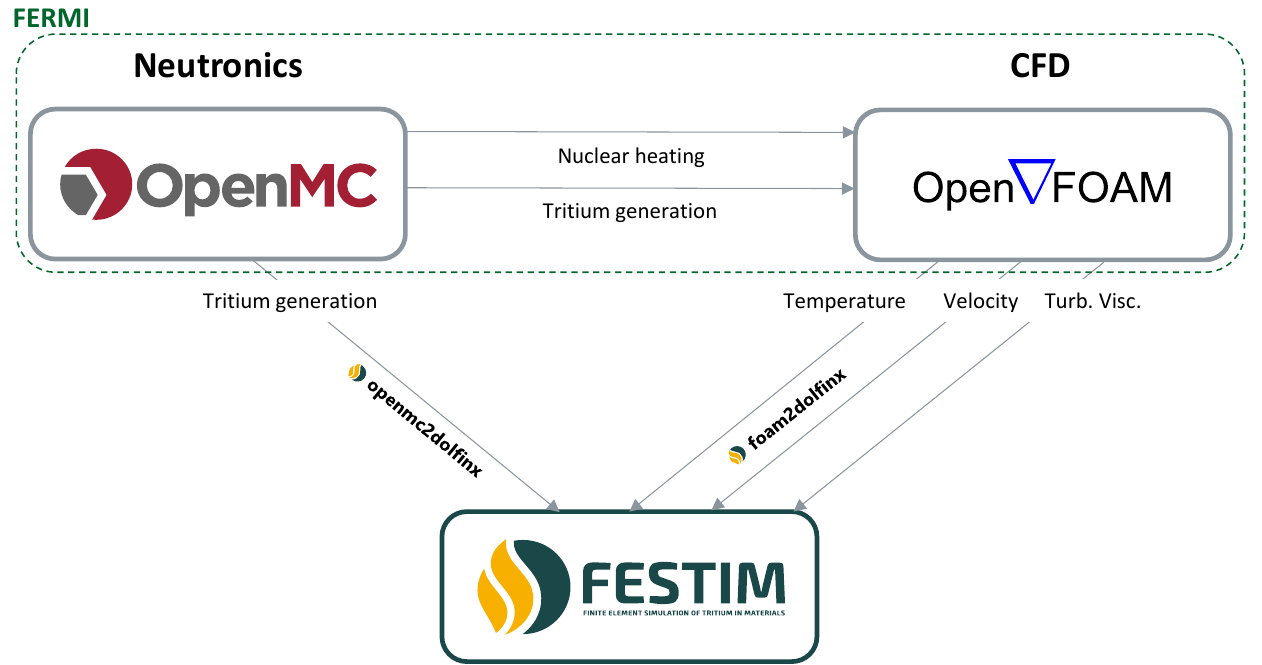}
    \caption{Coupling facilitated between OpenMC, OpenFOAM, and FESTIM using the \texttt{openmc2dolfinx} and \texttt{foam2dolfinx} packages. FERMI internal coupling is highlighted with the green dashed line.}
    \label{fig:multiphysics_coupling}
\end{figure}

The multiphysics modelling strategy adopted in this work builds on developments from the FERMI project~\cite{badalassi_fermi_2023}, which established a high-fidelity framework for coupled multiphysics simulations of fusion reactor blankets, including the liquid immersion blanket concept proposed for ARC-class tokamaks.
In particular, FERMI demonstrated detailed 3D OpenFOAM thermal hydraulic analyses informed by neutronics-derived volumetric heating under representative operating conditions. 
The present work leverages these established workflows and results, extending them to tritium transport modelling by integrating neutronics and CFD outputs into a finite element framework.
To enable this integration in a robust, reproducible manner, two lightweight Python packages are used to transfer spatially resolved fields from OpenMC and OpenFOAM into DOLFINx, which serves as the numerical backend for FESTIM; see figure~\ref{fig:multiphysics_coupling}.

The \texttt{openmc2dolfinx} package enables the direct import of OpenMC tally results, such as tritium generation rates or nuclear heating, from \texttt{.vtk} files. 
Because tally values are constant within each mesh cell, they can be mapped cell-by-cell onto an equivalent DOLFINx mesh without interpolation, preserving the fidelity of the neutronics solution. 
Both structured and unstructured tally meshes are supported, ensuring flexibility across different OpenMC simulations.

The \texttt{foam2dolfinx} package streamlines the import of OpenFOAM fields into DOLFINx. 
It reads \texttt{.foam} case files and reconstructs a matching finite element mesh by mapping OpenFOAM cell connectivity to DOLFINx element topology. 
This direct correspondence avoids interpolation overhead while ensuring accurate transfer of quantities such as velocity, temperature, or turbulent viscosity. 
Transient data can be imported at specified time steps, and fields can be restricted to subdomains if necessary.

Together, these tools provide a practical and efficient way to pass external fields into FESTIM. 
Neutronics simulations provide tritium source terms (and heat deposition by neutron and gamma heating for the CFD heat transfer calculations), while CFD models supply velocity, turbulent viscosity, and temperature fields; both are seamlessly integrated into a unified multiphysics workflow. 
This modular strategy, in line with the partitioned multiphysics approach of FERMI, ensures that each solver remains specialised in its domain while enabling a consistent combination of their outputs in the blanket modelling effort.

\subsection{Neutron transport using OpenMC}

%    \item Geometry/type of mesh
OpenMC utilised the DAGMC library \cite{wilson_acceleration_2010} for geometry modelling, in which volumes are represented by tessellated surfaces.
The geometry can be created by meshing an existing CAD file and labelling the volumes bound by the surfaces. 
The labels are then read in OpenMC, where each volume is treated as an OpenMC cell with its material assignment.
An initial uniform temperature of \SI{800}{K} was assumed for the FLiBe. 
The DAGMC geometry is prepared using FERMI, which leverages the Python API of Cubit \cite{blacker_cubit_2016}. 
Additionally, in OpenMC, volumetric tetrahedral meshes can be provided for tallies. 
In the same workflow, a volumetric mesh is generated from the same CAD file and provided to OpenMC for spatially resolved tritium generation and heat deposition tallies. 

The neutron source is modelled as a point cloud, with the space, energy, and angular distributions sampled using the tools from Fausser et al.~\cite{fausser_tokamak_2012}.
The ENDF/B-VII.1 \cite{chadwick_endfb-vii1_2011} nuclear data library was used. 
To ensure good statistics for the tetrahedral mesh tallies, a subroutine was set using the OpenMC Python API to monitor the maximum Monte Carlo relative error.
The subroutine was set to continue running until the highest relative error across the tetrahedral mesh tallies fell below \SI{1}{\%}. 
Nine hundred (900) million particles were simulated, with 300,000 particles per batch, using the OpenMC built-in Method of Automatic Generation of Importances by Calculation (MAGIC)~\cite{davis2011comparison} to update the weight windows of each batch for variance reduction.
The material assignments for each layer, not considering any impurities, are shown in Table \ref{tab:neutronics_properties}.

\begin{table}[H]
    \centering
    \caption{Densities used in neutronics model.}
    \begin{tabular}{L{0.4\linewidth} L{0.3\linewidth} R{0.15\linewidth}}
        Layer name & Material & Density\\
         & & (\si{g.cm^{-3}}) \\
        \hline
        First wall & Tungsten & 19.3 \\
        Inner vacuum vessel & Inconel 718 & 8.19 \\
        Coolant channel & FLiBe (90\% Li6) & 1.924\\
        Neutron multiplier & Be & 1.847 \\
        Outer vacuum vessel & Inconel 718 & 8.19 \\
        Bulk blanket & FLiBe (90\% Li6) & 1.924\\
    \end{tabular}
    \label{tab:neutronics_properties}
\end{table}

\subsection{Heat transfer and fluid dynamics using OpenFOAM}

The thermal hydraulic behaviour of the ARC blanket coolant was modelled using the open-source CFD framework OpenFOAM~\cite{noauthor_openfoam_nodate}, which provides a robust finite volume formulation for coupled momentum and heat transfer problems in complex geometries. 
OpenFOAM was selected for its established use in nuclear and fusion thermal hydraulics~\cite{de_pietri_development_2023, mistrangelo_determination_2018, caravello_openfoam_2024}, its flexibility for incorporating volumetric source terms, and its suitability for integration within a broader multiphysics workflow~\cite{seo_review_2021}.

The simulations were performed using a custom transient solver, \texttt{rhoBlanketFoam}, derived from the standard PIMPLE algorithm, and using a spatially resolved heat deposition field mapped from OpenMC neutronics.
This custom solver was developed as part of the FERMI project.
The solver advances the conservation equations for mass, momentum, and energy for a single-phase fluid.
Although the equations are formulated in a mildly compressible form, the density was treated as constant, resulting in an effectively incompressible flow.
The solved fields include velocity, pressure, temperature, and density. 
Turbulence was modelled using a Reynolds-averaged Navier–Stokes approach with the $k$-$\omega$ SST closure, including standard wall functions for near-wall treatment. 
Gravity was set to zero, and no buoyancy effects were included. Hybrid first- and second-order spatial discretisation schemes are used for flux calculations in the momentum and energy equations, and pure second-order schemes are used for derivatives, with non-orthogonal correctors to account for mesh non-orthogonality, given the arbitrary blanket profile.

\begin{figure}[H]
    \centering
    % trim={<left> <lower> <right> <upper>}
    \begin{subfigure}[c]{0.38\linewidth}
        \centering
        \includegraphics[width=\linewidth, trim={11cm 4cm 11cm 4cm}, clip]{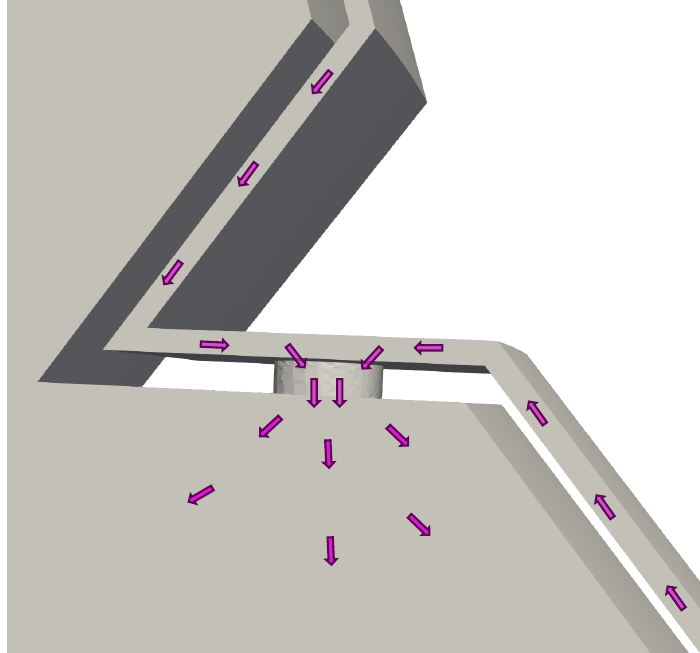}
        \caption{Interconnect region}
        \label{fig:ARC_blanket_interconnect}
    \end{subfigure} 
    \begin{subfigure}[c]{0.50\linewidth}
        \centering
        \includegraphics[width=\linewidth, trim={10cm 0cm 7cm 0cm}, clip]{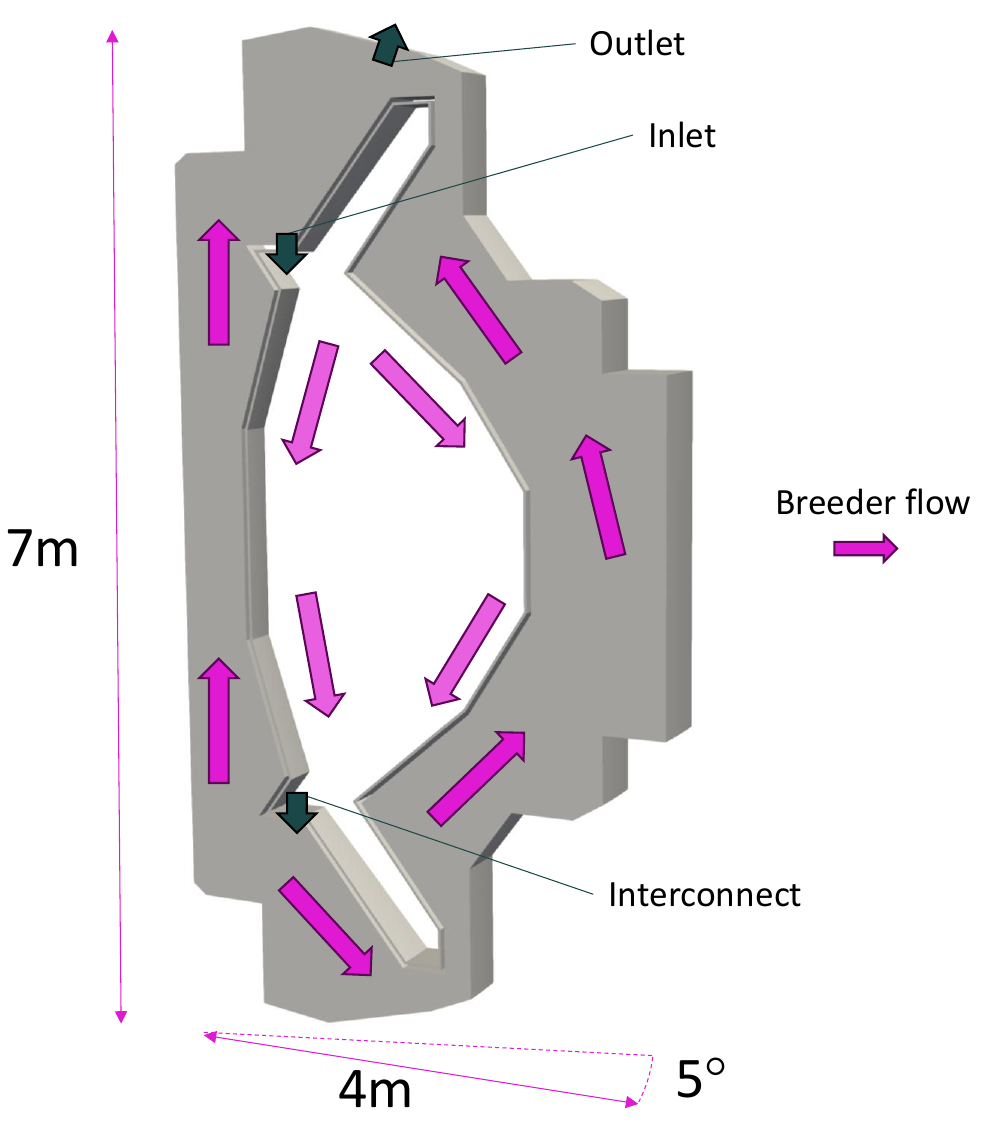}
        \caption{Breeder flow around blanket}
        \label{fig:breeder_flow}
    \end{subfigure}
    \caption{Visual representation of the breeder flow path in the ARC model, outlining the dimensions and locations of the inlet, outlet, and interconnect between the inner and outer sections of the blanket.}
    \label{fig:arc_geom}
\end{figure}

Thermal transport was formulated in terms of sensible enthalpy, assuming constant thermophysical properties; see Table \ref{tab:thermophysical_properties}. 
The volumetric heat generation term, representing internal heating within the fluid domain, was obtained from OpenMC neutronics simulations and mapped using tools in the FERMI workflow. 
Turbulent heat transfer was accounted for through a turbulent thermal diffusivity with a constant turbulent Prandtl number.

\begin{table}[H]
    \centering
    \caption{Thermophysical properties of FLiBe used in the CFD model.}
    \begin{tabular}{L{0.45\linewidth} R{0.1\linewidth} L{0.17\linewidth}}
        Parameter & Value & Units \\
        \hline
        Density, $\rho$ & 1940 & \si{kg.m^{-3}} \\
        Specific heat capacity, $c_p$ & 2400 & \si{J.kg^{-1}.K^{-1}} \\
        Dynamic viscosity, $\mu$ & 0.006 & \si{Pa.s} \\
        Prandtl number, Pr & 14.4 & -- \\
        Turbulent Prandtl number, Pr$_t$ & 0.85 & --
    \end{tabular}
    \label{tab:thermophysical_properties}
\end{table}

The computational domain represents a reduced \SI{5}{\degree} sector of the blanket coolant channel (see Figure~\ref{fig:arc_geom}), exploiting geometric symmetry where applicable.
A symmetry condition was imposed on the sector's side walls, consistent with the assumption of perfect geometric and flow symmetry among the 72 identical sectors that make up the full blanket.
Velocity boundary conditions consisted of a fixed inlet velocity, no-slip walls, a zero-gradient outlet, and symmetry conditions at the toroidal faces; see Table \ref{tab:bc_values}.
Pressure was prescribed at the outlet, with zero-gradient conditions elsewhere.
The temperature boundary conditions included a fixed inlet temperature, prescribed temperatures at selected solid–fluid interfaces, adiabatic conditions on the remaining walls, and zero-gradient conditions at the outlet. 
Turbulent quantities were specified at the inlet to correspond to \SI{20}{\%} turbulence intensity and treated with wall functions at solid boundaries.

\begin{table*}[t]
    \centering
    \caption{Imposed boundary condition values used in the CFD model.}
    \begin{tabular}{L{0.1\linewidth} L{0.40\linewidth} R{0.05\linewidth} L{0.05\linewidth}}
        Parameter & Description & Value & Units \\
        \hline
        $\boldsymbol{U}_{\mathrm{in}}$ & Inlet velocity & 2 & \si{m.s^{-1}} \\
        $T_{\mathrm{in}}$ & Inlet temperature & 800 & \si{K} \\
        $p_{\mathrm{out}}$ & Outlet pressure & 0 & \si{Pa} \\
        $T_{\mathrm{wall, Inc}}$ & Wall temperature at Inconel interfaces & 900 & \si{K} \\
        $T_{\mathrm{wall, Be}}$ & Wall temperature at beryllium interfaces & 850 & \si{K} \\
        $T_{\mathrm{wall, VV}}$ & Wall temperature at vacuum vessel interfaces & 1000 & \si{K} \\
        $k_{\mathrm{in}}$ & Inlet turbulent kinetic energy & 0.12 & \si{m^2.s^{-2}} \\
        $\omega_{\mathrm{in}}$ & Inlet specific dissipation rate & 0.6 & \si{s^{-1}} \\
        $All$ & Symmetry at toroidal faces & &
    \end{tabular}
    \label{tab:bc_values}
\end{table*}

This modelling approach captures the dominant forced convection heat transfer behaviour in the blanket coolant but neglects density variations, buoyancy effects, and multiphase phenomena.
The use of Reynolds-averaged turbulence modelling limits the resolution of transient and small-scale flow structures.
In addition, volumetric source terms were prescribed based on the heat deposition from neutronics results.
These assumptions are appropriate for the present study, in which fluid mechanics primarily provides temperature and flow fields for subsequent multiphysics analyses rather than being the primary focus.
The mapping of neutronics heat deposition to the OpenFOAM mesh is handled within FERMI through \texttt{.vtk} files. 
A Python converter tool was developed, building on the \\ \texttt{vtkUnstructuredToFoam} capability and using \texttt{mapFields} within OpenFOAM to interpolate between OpenMC and OpenFOAM meshes \cite{bae_impact_2026}. 
The nearest-neighbour mapping method was found to conserve global energy between meshes to within \SI{0.1}{\%}.

\begin{table*}[b]
    \centering
    \caption{Tritium transport parameters.}
    \begin{tabular}{L{0.1\linewidth} L{0.4\linewidth} R{0.1\linewidth} L{0.05\linewidth}}
        Parameter & Description & Value & Units \\
        \hline
        $D_{\mathrm{0}}$ & Fickian diffusion pre-exponential factor & $6.75\times10^{-6}$ & \si{m^{2}.s^{-1}} \\
        $E_{\mathrm{D}}$ & Fickian diffusion activation energy & $0.615$ & \si{eV} \\
        $c_{\mathrm{in}}$ & Inlet concentration & 0 & \si{m^{-3}} \\
        $\mathrm{Sc}_t$ & Turbulent Schmidt number & 0.5 & -- \\
        $\delta$ & Numerical tuning parameter & 0.1 & -- \\
    \end{tabular}
    \label{tab:trit_parameters}
\end{table*}

\subsection{Tritium transport}
    \label{subsec:HyT-meth}

Tritium transport within the ARC breeding blanket is governed by the combined effects of diffusion, bulk advection due to molten salt flow, and volumetric generation from nuclear reactions.
In the liquid immersion blanket concept, tritium produced in the FLiBe breeder is transported by the coolant before being extracted or interacting with surrounding structures.
Capturing this behaviour, therefore, requires a transport model that accounts for spatially varying source terms, temperature-dependent material properties, and mixing induced by the coolant flow, while remaining compatible with the multiphysics workflow described above.

Under these assumptions, tritium transport in the breeder is modelled using a macroscopic advection–diffusion formulation, which is well suited to continuum-scale simulations of molten salts.

\begin{equation}
    \frac{\partial \cm}{\partial t} = \nabla\cdot(D \nabla \cm) + S - \nabla \cdot (\mathbf{u} \cm)
    \label{tt_1}
\end{equation}

Equation \eqref{tt_1} describes the rate equation used to model tritium transport, with a coupling advection term, $\nabla \cdot (\mathbf{u} \cm)$, and a diffusion term, $\nabla\cdot \left(D\nabla c_\mathrm{m}\right)$, to describe the diffusion of mobile hydrogen particle concentrations, $c_\mathrm{m}$, in units of m$^{-3}$.
$D = D_{0} \cdot \exp(-E_{\text{D}}/(k_{B}\, T))$ is the Fickian diffusion coefficient of interstitial tritium in units of $\si{m^{2}.s^{-1}}$, $k_{B}$ = \SI{8.617e-5}{eV.K^{-1}} is the Boltzmann constant, and $S$ is the volumetric source term of mobile tritium in units of $\si{m^{-3}.s^{-1}}$.
The mean molecular diffusivity of tritium in FLiBe was evaluated from a hydrogen transport properties database (HTM)~\cite{delaporte-mathurin_remdelaportemathurinh-transport-materials_2024}; see Table \ref{tab:trit_parameters}.
The Soret effect is not considered in this work.
Hydrogen trapping is not considered in this work, as only the liquid breeder is present.
The source term is obtained from the tritium production tally generated by OpenMC and ported to DOLFINx for use in the formulation via the \texttt{openmc2dolfinx} package.
$\mathbf{u}$ represents the velocity field of the liquid FLiBe breeder material in units of \si{m.s^{-1}}, and similarly to the source term, is ported to DOLFINx from OpenFOAM using the package \texttt{foam2dolfinx}.

\begin{figure*}[b]
    \centering
    % trim={<left> <lower> <right> <upper>}
    \begin{subfigure}{0.32\linewidth}
        \centering
        \includegraphics[width=\linewidth]{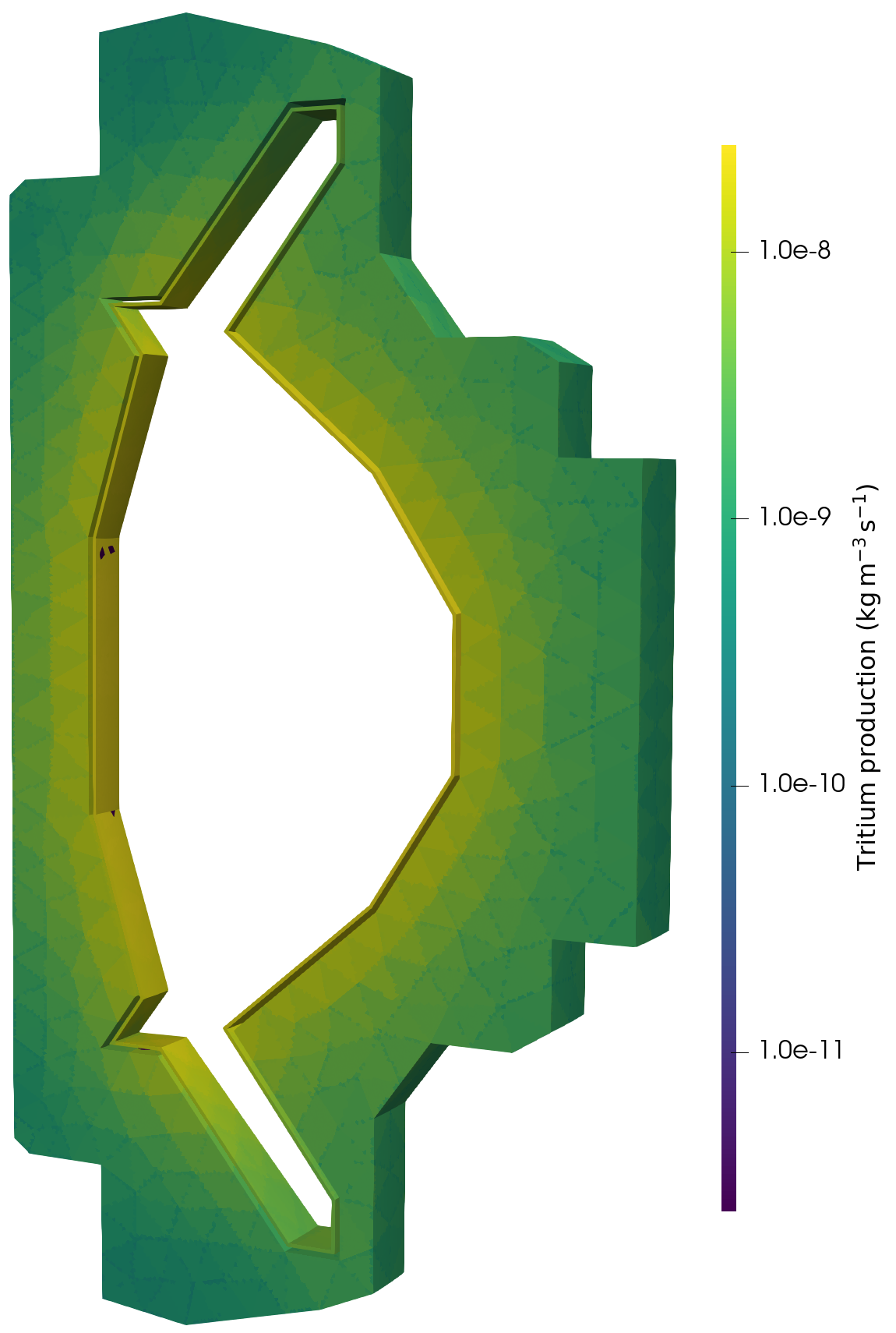}
        \caption{Tritium generation}
        \label{fig:res_mc_t_gen}
    \end{subfigure} 
    \quad
    \begin{subfigure}{0.32\linewidth}
        \centering
        \includegraphics[width=\linewidth]{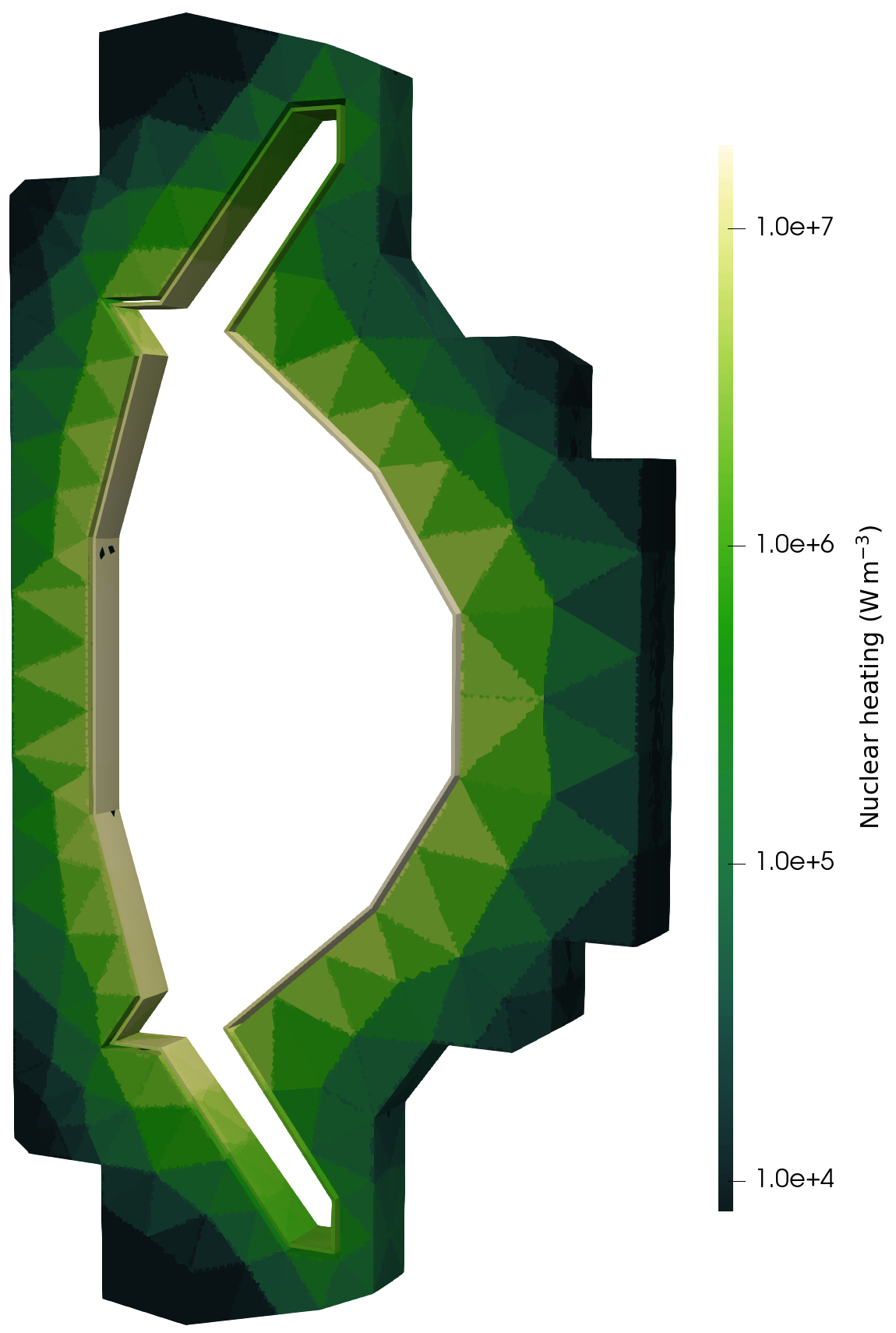}
        \caption{Nuclear heating}
        \label{fig:res_mc_nuc_heating}
    \end{subfigure}
    \caption{OpenMC results.}
    \label{fig:res_mc}
\end{figure*}

Turbulence in the breeder flow enhances tritium mixing and transport well beyond what molecular diffusion alone can achieve. 
To account for this effect, the tritium transport model includes a \textit{turbulent diffusion} term by defining an effective diffusivity:
\begin{equation}
D_{\text{turb}} = \frac{\nu_t}{\mathrm{Sc}_t},
\end{equation}
where $D_{\text{turb}}$ is the turbulent diffusivity of hydrogen, $\nu_t$ is the turbulent kinematic viscosity (obtained from the fluid dynamics simulation), and $\mathrm{Sc}_t$ is the turbulent Schmidt number.
The turbulent Schmidt number, $\mathrm{Sc}_t$, characterises the ratio between turbulent momentum diffusivity and mass diffusivity. 
For molten salt coolants such as FLiBe, experimentally validated values of the turbulent Schmidt number are scarce; in the absence of material-specific data, the adopted value falls within the commonly assumed range for turbulent mass transport in liquids and reflects enhanced scalar mixing relative to momentum diffusion.
This formulation enables the hydrogen transport model to respond dynamically to local turbulence intensity, capturing enhanced tritium transport in regions of high shear or recirculating flow.

To ensure numerical stability when solving the advection-diffusion equation at high Péclet numbers, a stabilisation method was also employed. 
Solving advection-dominated transport problems with the finite element method (FEM) can result in oscillations unless additional stabilisation terms are introduced.
A consistent stabilisation strategy was adopted, analogous to streamline upwind Petrov–Galerkin (SUPG) or artificial diffusion methods~\cite{comsol_understanding_2020}, which introduce additional diffusion aligned with the direction of flow to preserve physical accuracy without overly damping the solution. 
This improves the solver's convergence, particularly in regions with strong gradients or recirculating flows where advection dominates.
This is implemented by
\begin{equation}
    D_{\mathrm{art}} = \delta h \|\mathbf{u}\|,
    \label{eq:stabliser}
\end{equation}
where $D_{\mathrm{art}}$ is the artificial diffusion coefficient, $\delta$ is a dimensionless tuning parameter, $h$ is the mesh element size in \si{m}, and $\|\mathbf{u}\|$ is the magnitude of the velocity.
This value is used in tandem with the turbulent diffusivity term to evaluate the effective diffusivity term, $D_{\mathrm{eff}}$ given by equation~\eqref{eq:deff}, which is then used in place of the standard hydrogen diffusivity, $D$; see equation~\eqref{tt_1}.
\begin{equation}
    D_{\mathrm{eff}} = D + D_{\mathrm{turb}} + D_{\mathrm{art}}
    \label{eq:deff}
\end{equation}

A homogeneous Dirichlet boundary condition is applied to the inlet, which assumes a perfect purification system.

The FESTIM code~\cite{dark_festim_2026} was employed and extended to model tritium transport in the ARC breeding blanket, solving equation~\eqref{tt_1}. 
It uses the FEM to simulate diffusion and trapping of hydrogen isotopes based on macroscopic rate equations~\cite{hodille_macroscopic_2015}.

\subsubsection{OpenFOAM passive scalar transport formulation}

A code-to-code comparison was performed against an equivalent passive scalar transport solver implemented in OpenFOAM~\cite{sircar_2024}, which simulates the transport of mobile tritium analogously to equation~\eqref{tt_1},
\begin{align} \label{eq:cm_of}
    \frac{\partial \cm}{\partial t} + \nabla \cdot (\mathbf{u}\cm) &= - \nabla \cdot (J) + S \\
    J &= -D\nabla{\cm} - \frac{\nu_t}{\mathrm{Sc}_t}\nabla{\cm}, \nonumber
\end{align}
where $J$ denotes the tritium flux.
The main difference between OpenFOAM and FESTIM is that the former is a finite-volume-based code, whereas the latter is a finite-element-based code.
The discretisation of the advection term is handled in a conservative form in OpenFOAM, as shown in equation~\eqref{eq:cm_of}.
The exact setup of FESTIM---including initial conditions, boundary conditions, and neutronics generation rates (coupled using the approaches described in \cite{badalassi_fermi_2023})---was reproduced in OpenFOAM to solve the equations.
However, there are some differences in the numerics due to the different discretisation methods of the two codes.
To solve tritium transport in OpenFOAM, a backwards Euler (second-order) time discretisation scheme was used along with the \textit{limitedLinear} scheme for discretising the advective term, with all other terms discretised using the \textit{Gauss linear} scheme, which is a second-order scheme. 
The \textit{limitedLinear} scheme is a hybrid scheme that is primarily second-order but relaxes to first-order in regions of high gradients. 
In OpenFOAM, the tritium concentration is solved in units of $\si{kg.m^{-3}}$.

\begin{figure*}[b]
    \centering
    \begin{subfigure}{0.32\linewidth}
        \centering
        \includegraphics[width=\linewidth]{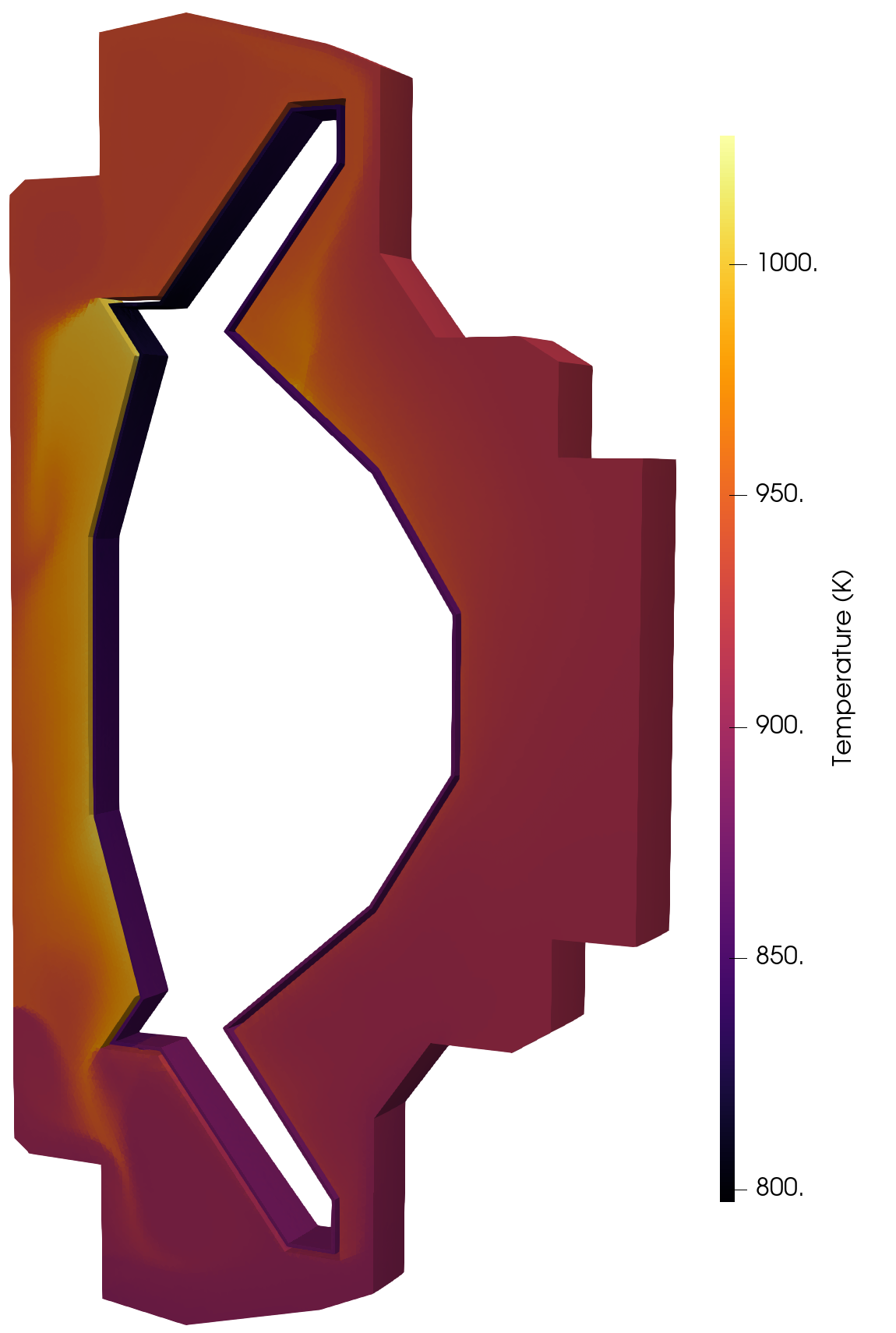}
        \caption{Temperature}
        \label{fig:res_foam_T}
    \end{subfigure}
    \begin{subfigure}{0.32\linewidth}
        \centering
        \includegraphics[width=\linewidth]{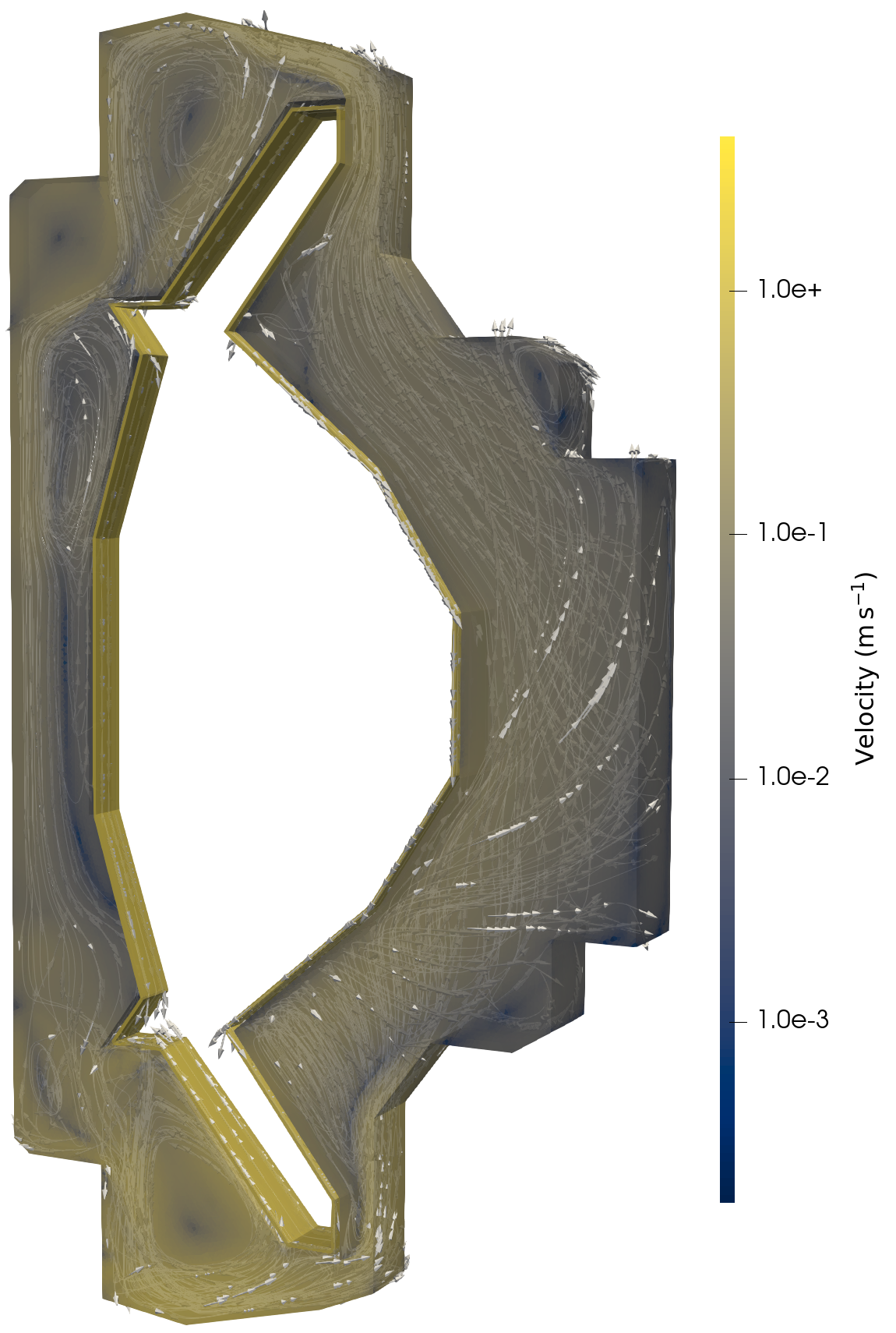}
        \caption{Velocity}
        \label{fig:res_foam_u}
    \end{subfigure}
    \begin{subfigure}{0.32\linewidth}
        \centering
        \includegraphics[width=\linewidth]{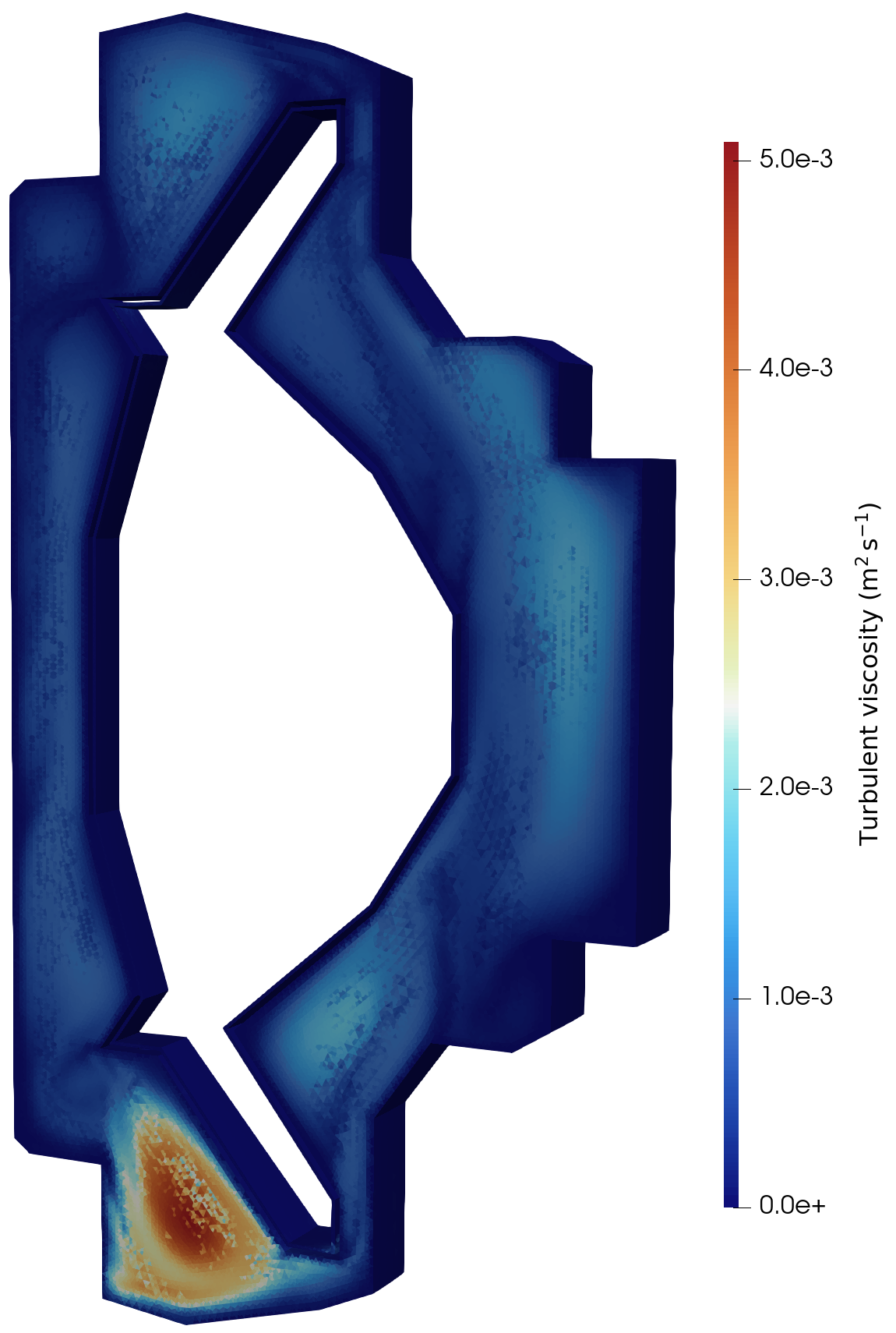}
        \caption{Turbulent viscosity}
        \label{fig:res_foam_nut}
    \end{subfigure}
    \caption{OpenFOAM results.}
    \label{fig:res_foam}
\end{figure*}

\section{Results}
All results presented in this section correspond to values extrapolated to the full breeding blanket.
The underlying transport model represents a \SI{5}{\degree} toroidal sector of the blanket, assuming perfect geometric and flow symmetry.
Under this assumption, the full blanket is composed of 72 identical sectors, each with its own inlet and outlet.
Tritium inventories, outlet fluxes, and outlet-averaged concentrations, for instance, are therefore obtained by scaling the results from the simulated sector to the full blanket.
All plots and figures in this work were generated with Matplotlib~\cite{hunter_j_d_matplotlib_2007} and ParaView~\cite{ayachit_paraview_2015}.

\subsection{OpenMC and OpenFOAM results}
    \label{subsec:res-neut}

The OpenMC results (figure \ref{fig:res_mc}) show a dramatic reduction with increasing distance from the neutron source (i.e., the plasma).
The tritium generation rate and heat deposition values are converted to SI units and normalised by the volume of each mesh element to obtain volumetric source terms.

In the OpenFOAM simulations, the flow velocities in the cooling channel are much higher than in the blanket due to the channel's much smaller cross-sectional area; flow velocities in the channel reach a maximum of \SI{12}{m.s^{-1}} compared with \SI{2}{m.s^{-1}} in the blanket. 
Although the channel lies closer to the plasma and therefore experiences much more intense nuclear heating (see figures~\ref{fig:res_mc_nuc_heating} and~\ref{fig:res_foam_T}), this high velocity, together with the fixed inlet temperature of the incoming FLiBe, provides efficient heat removal, so the channel stays comparatively cool, with temperatures ranging from approximately \SI{800}{K} at the inlet to \SI{870}{K}. 
In the blanket, by contrast, the much slower flow removes heat less effectively, and temperatures peak at around \SI{1000}{K}.

The flow field shows a region of very high mixing when the FLiBe flows through the channel and interconnects with the blanket.
The behaviour is similar to that of a jet, since the fast-moving molten salt in the channel enters a region of slowly moving fluid in the blanket.
This is potentially a region of high pressure drop and warrants further investigation in the ARC design.
As the flow moves upwards, it carries heat and tritium produced by neutron interactions with the molten salt. 
Towards the top, given that the top divertor creates a flow restriction near the outlet, a recirculation zone is observed (see figures \ref{fig:res_foam_u} and~\ref{fig:zoom_vel}). 
These flow features have important implications for tritium transport within the blanket, as discussed next.

\subsection{FESTIM results}
\label{subsec:res-festim}

\begin{figure}[t]
    \centering
    \includegraphics[width=\linewidth]{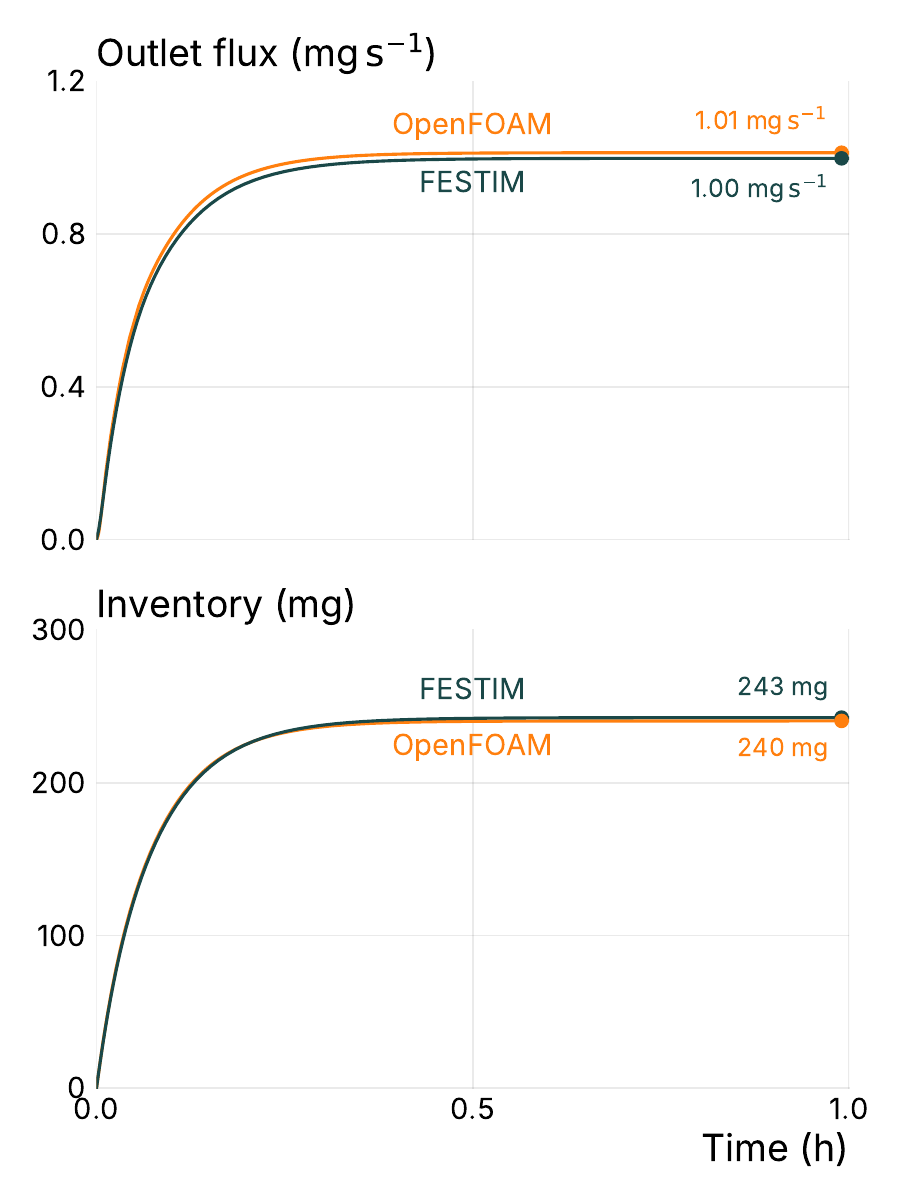}
    \caption{Temporal evolution of the total tritium inventory (bottom) and the mass flux of tritium leaving the outlet (top) in the ARC blanket model, comparing the FESTIM and OpenFOAM solutions.}
    \label{fig:plot_transient}
\end{figure}

Under steady-state conditions, the total tritium inventory within the liquid breeder is predicted to be approximately \SI{243}{mg}.
Peak local tritium concentrations reach values on the order of \SI{2.5}{mg.m^{-3}}.
At the blanket outlet, a steady-state tritium mass flux of approximately \SI{1.0}{mg.s^{-1}} is obtained, corresponding to an outlet-averaged tritium concentration of approximately \SI{0.90}{mg.m^{-3}}.
The blanket inventory and outlet flux approach steady state within approximately \SI{30}{min} (figure~\ref{fig:plot_transient}). 
This build-up time is of a similar order to the characteristic blanket residence times assumed in system-level analyses of the ARC fuel cycle~\cite{meschini_modeling_2023}. 
The predicted inventory and build-up time differ from earlier ARC-relevant estimates, as expected given differences in geometry, dimensionality, and modelling assumptions across studies~\cite{ferrero_preliminary_2022}.

\begin{figure}[t]
    \centering
    \begin{subfigure}{0.7\linewidth}
        \centering
        \includegraphics[width=\linewidth]{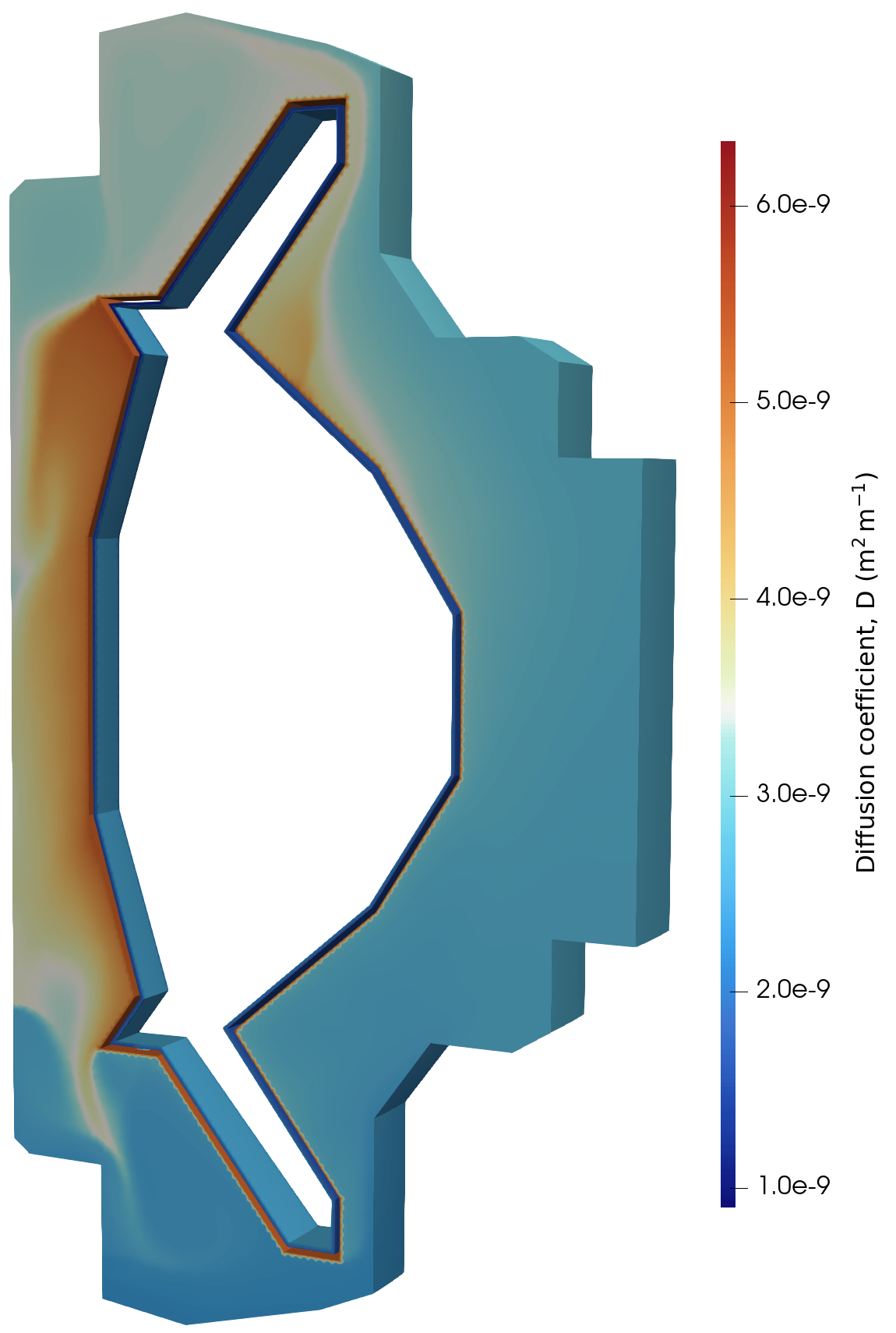}
        \caption{Fickian diffusivity}
        \label{fig:fick_diffusion_field}
    \end{subfigure} \\ 
    \begin{subfigure}{0.7\linewidth}
        \centering
        \includegraphics[width=\linewidth]{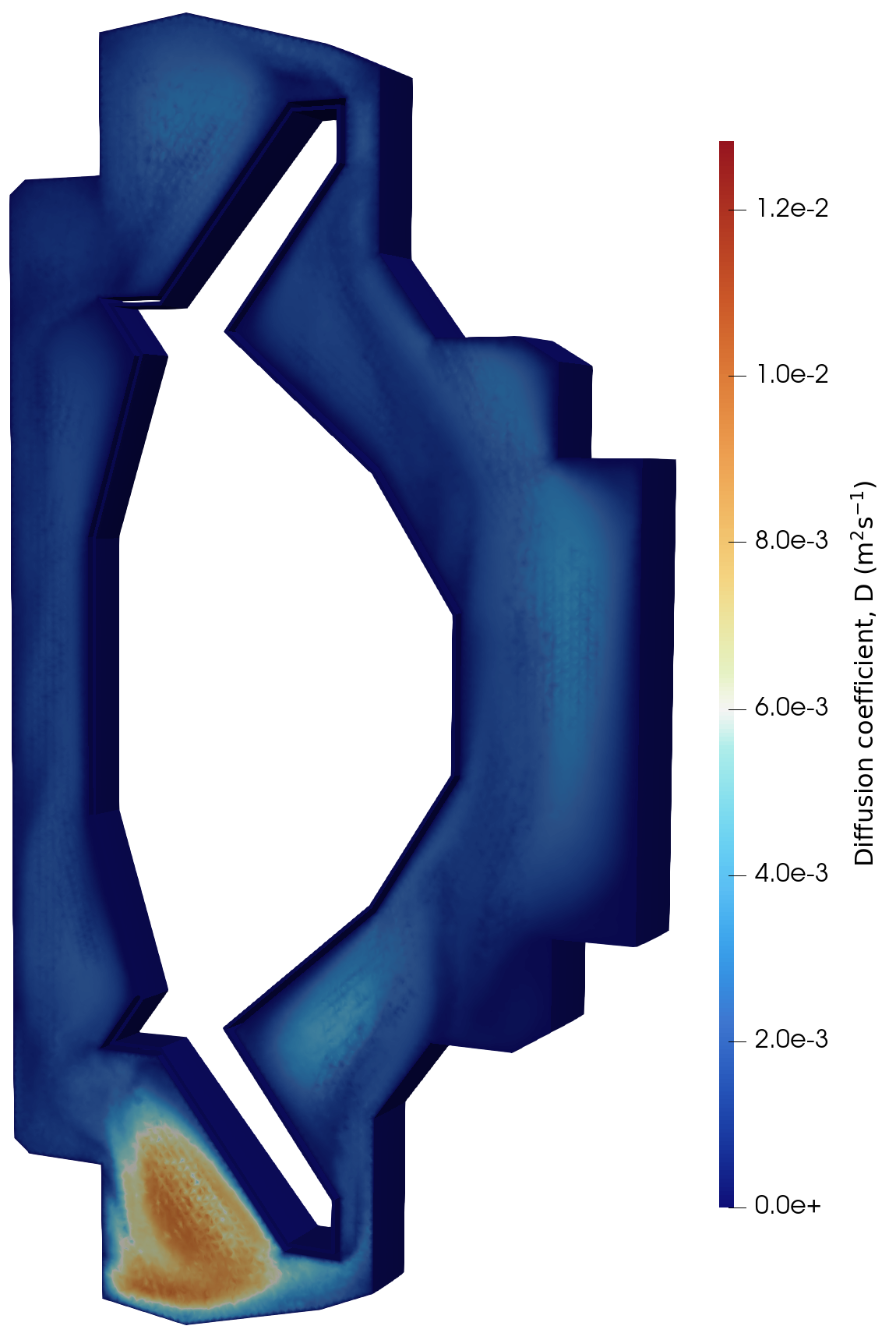}
        \caption{Effective diffusivity}
        \label{fig:diffusion_field}
    \end{subfigure}
    \caption{Diffusivity fields}
    \label{fig:res_diffusion}
\end{figure}

It should be emphasised that these values represent a lower-bound estimate. 
The homogeneous Dirichlet condition imposed at the inlet corresponds to an idealised, perfectly efficient extraction system; in practice, a finite inlet concentration would raise the equilibrium concentration throughout the breeder and hence increase the total inventory and outlet-averaged concentration.
The reported inventory should therefore be interpreted as a conservative minimum for the assumed geometry and operating conditions.

\begin{figure*}[t]
    \centering
    % Left block: 2x2
    \begin{minipage}[c]{0.58\linewidth}
        \centering
        \begin{subfigure}{0.48\linewidth}
            \centering
            \includegraphics[width=\linewidth]{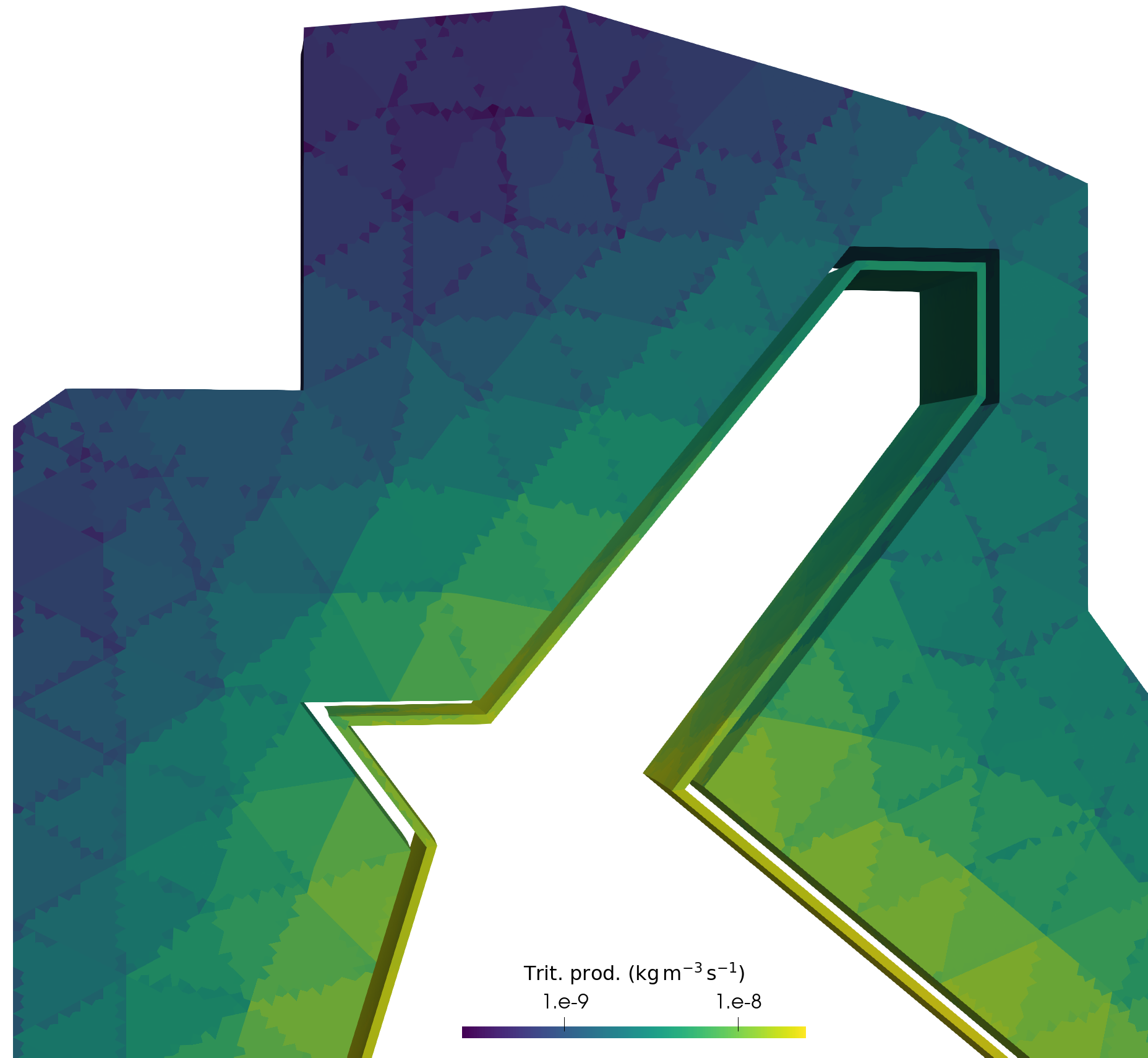}
            \caption{Tritium production}
            \label{fig:zoom_t_prod}
        \end{subfigure}
        \hfill
        \begin{subfigure}{0.48\linewidth}
            \centering
            \includegraphics[width=\linewidth]{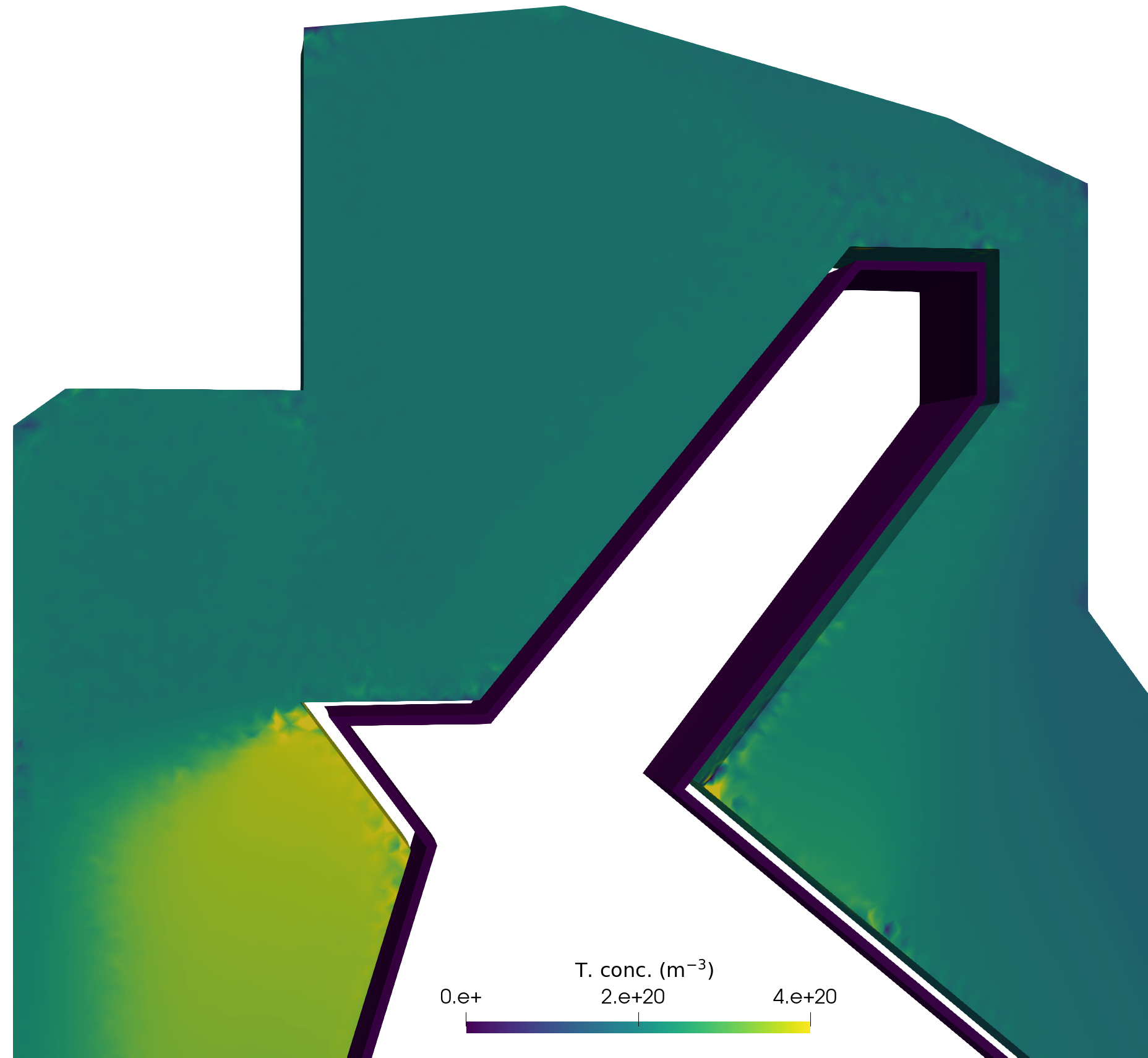}
            \caption{Tritium concentration}
            \label{fig:zoom_mobile_conc}
        \end{subfigure}
        \vspace{0.8em}
        \begin{subfigure}{0.48\linewidth}
            \centering
            \includegraphics[width=\linewidth]{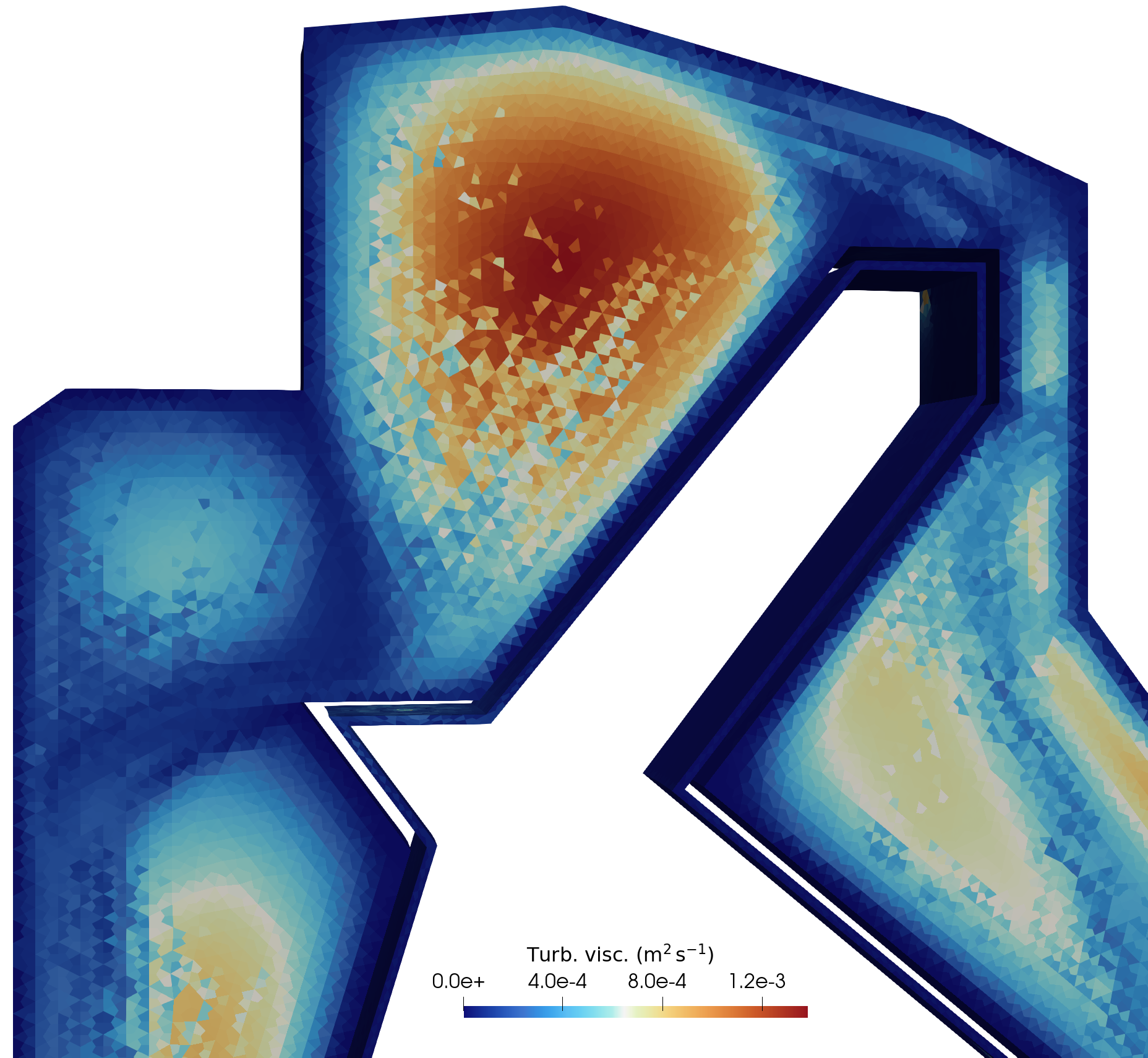}
            \caption{Turbulent viscosity}
            \label{fig:zoom_turb_visc}
        \end{subfigure}
        \hfill
        \begin{subfigure}{0.48\linewidth}
            \centering
            \includegraphics[width=\linewidth]{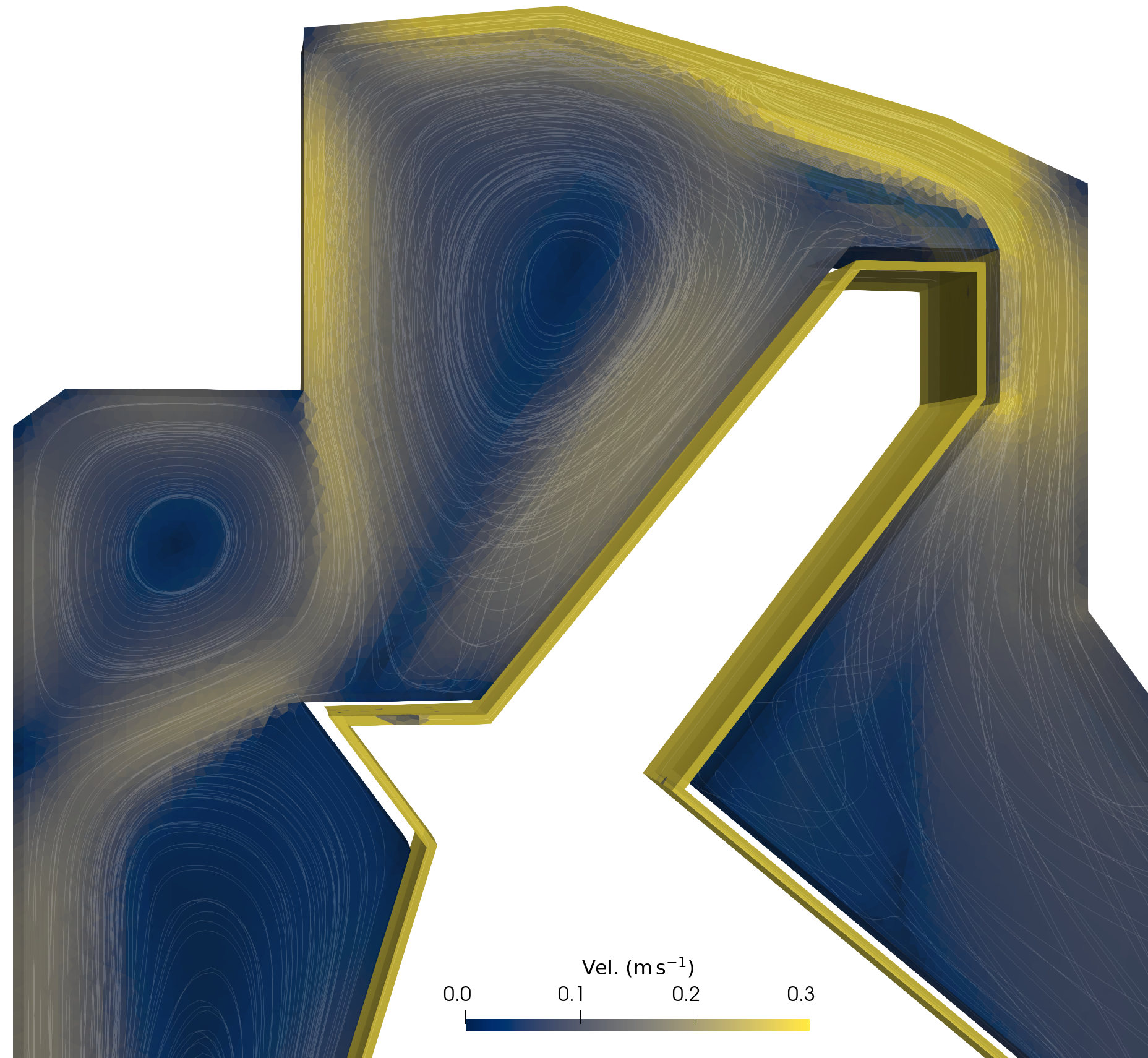}
            \caption{Velocity}
            \label{fig:zoom_vel}
        \end{subfigure}
    \end{minipage}
    \hfill
    % Right block: single figure, vertically centred relative to the left block
    \begin{minipage}[c]{0.38\linewidth}
        \centering
        \begin{subfigure}{\linewidth}
            \centering
            \includegraphics[width=\linewidth]{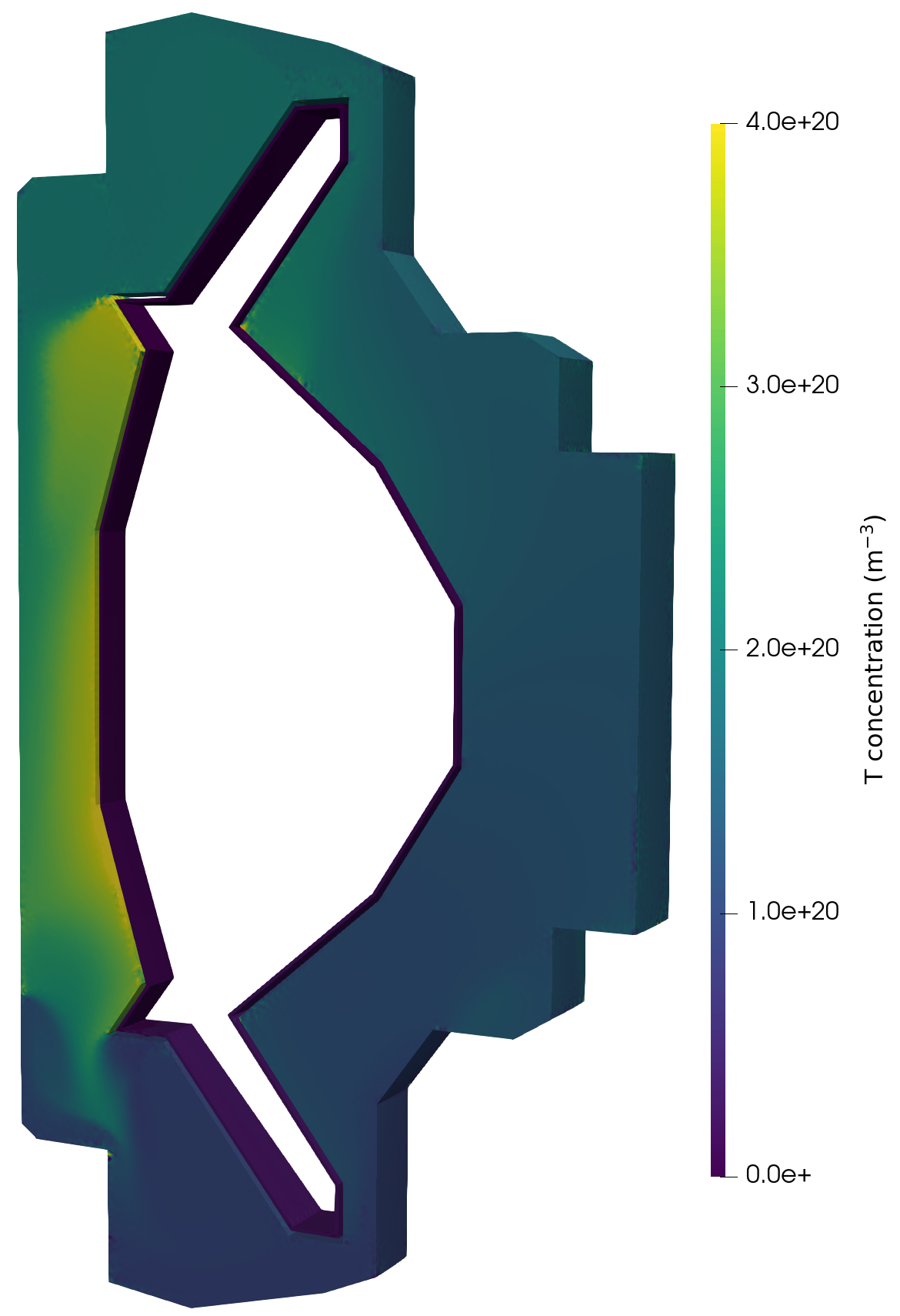}
            \caption{Steady-state tritium concentration}
            \label{fig:res_festim_mobile}
        \end{subfigure}
    \end{minipage}
    \caption{Multiphysics fields within the upper region of the blanket sector. The zoomed views highlight the interplay between local tritium production, flow structure, turbulence, and geometry, which together govern the resulting spatial distribution of tritium concentration.}
    \label{fig:zoom}
\end{figure*}

\begin{figure}[t]
    \centering
    \includegraphics[width=\linewidth]{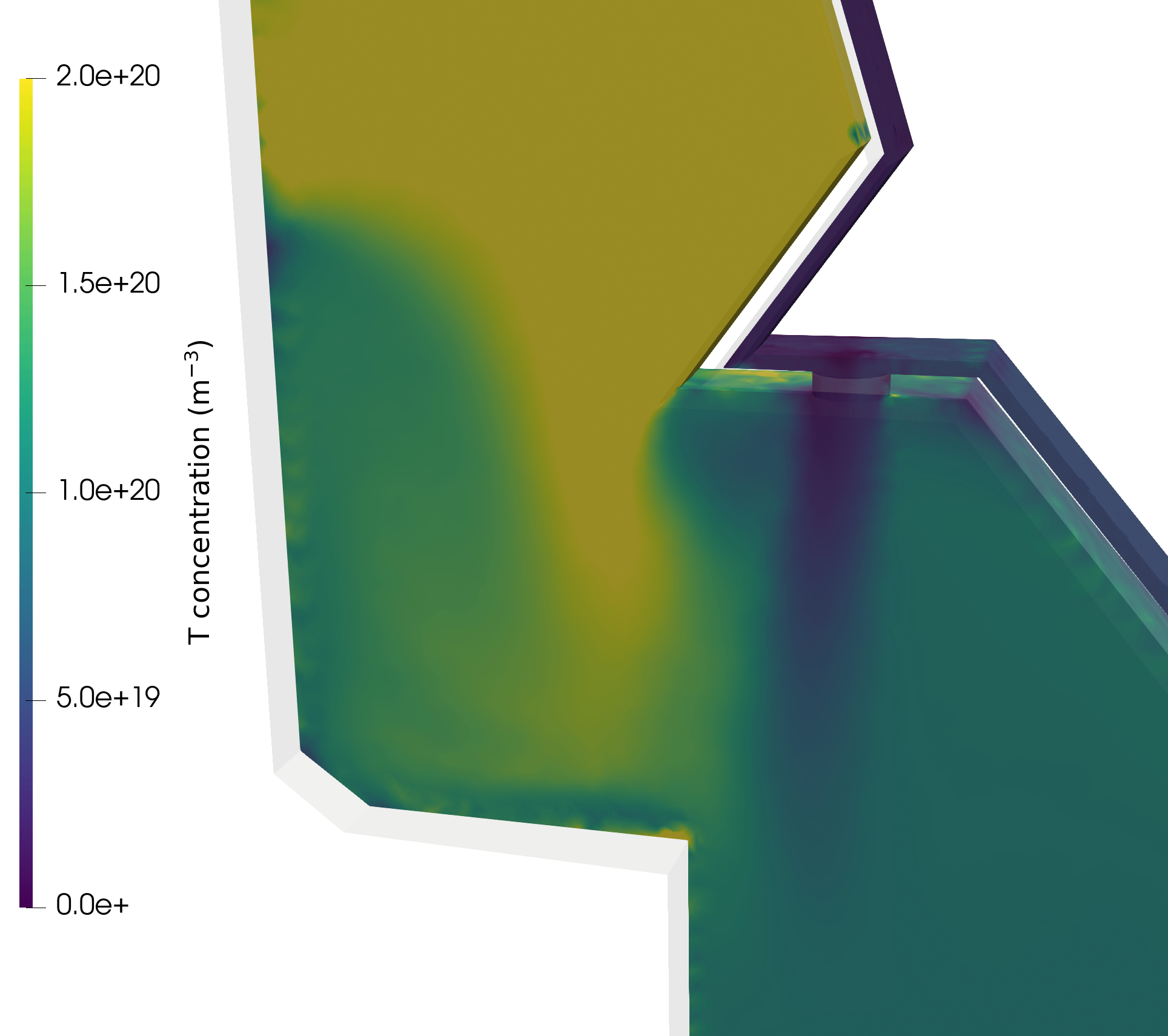}
    \caption{Tritium concentration in the vicinity of the interconnect, where the fast-moving FLiBe from the coolant channel enters the slower-moving bulk blanket. The colour scale is saturated to emphasise the structure of the low-concentration jet issuing from the channel.}
    \label{fig:conc_interconnect}
\end{figure}

The distribution of tritium concentration exhibits pronounced localisation effects, as shown in figure~\ref{fig:res_festim_mobile}.
Regions of elevated tritium concentration are strongly correlated with flow stagnation zones, indicating a strong sensitivity of tritium accumulation to the local flow structure within the breeder.
In these regions, reduced convective transport leads to localised concentration hot spots.
Conversely, areas characterised by high turbulence intensity exhibit enhanced effective diffusion and correspondingly lower tritium concentrations.
This behaviour is consistent with the dominant contribution of turbulence-induced diffusion to tritium transport: comparison of the Fickian and effective diffusivity fields (figure~\ref{fig:res_diffusion}) shows the latter exceeding the former by several orders of magnitude throughout the breeder.
In such regions, increased turbulent mixing dissipates concentration gradients, thereby limiting local tritium accumulation.
The interconnect region exhibits the converse effect, as shown in figure~\ref{fig:conc_interconnect}.
FLiBe issuing from the coolant channel into the bulk blanket forms a jet of comparatively low tritium concentration, despite the volumetric tritium generation rate being highest in the channel owing to its proximity to the plasma (figure~\ref{fig:res_mc_t_gen}).
This apparent contradiction is explained by residence time: the flow velocity in the channel reaches \SI{12}{m.s^{-1}}, an order of magnitude greater than in the blanket, so the salt traverses the high-generation region too rapidly for appreciable tritium to accumulate before entering the blanket.
Combined with the homogeneous Dirichlet condition at the inlet, this produces a persistent low-concentration plume that penetrates well into the blanket before being dispersed by turbulent mixing in the shear layer at the jet boundary.

Concentration in the breeder is therefore governed by the interplay between local production and local residence time, rather than by the production rate alone.
These results show that spatially resolved tritium transport simulations can be used to identify unfavourable flow features within the blanket.
In particular, regions prone to stagnation or weak mixing may be targeted for geometric optimisation to enhance convective or turbulent transport.

The tritium mass flux at the blanket outlet provides a useful system-level metric for assessing overall blanket performance and tritium throughput.
From the transient simulations, an outlet tritium mass flux of approximately \SI{1.0}{mg.s^{-1}} is obtained under steady-state conditions.
However, the transient evolution of the outlet flux and blanket inventory, shown in figure~\ref{fig:plot_transient}, provides additional insight beyond the steady-state value alone.
The results indicate that the outlet tritium flux approaches its steady-state value over approximately \SI{30}{min}.
This timescale is significant when compared with expected plasma pulse durations in fusion devices.

For reference, pulse lengths of order \SI{30}{min} are anticipated for ITER~\cite{shimada_chapter_2007}, while no definitive pulse length has yet been published for ARC-class devices.
Although ARC-class reactors are generally envisaged to operate with long or quasi-steady-state plasma conditions~\cite{sorbom_arc_2015}, the predicted transient response time remains highly relevant.
If the time required for the blanket to reach steady-state tritium throughput exceeds a single plasma pulse, multiple pulses may be needed for the blanket inventory and outlet flux to approach steady state.
In such scenarios, the time for the blanket inventory and outlet flux to reach steady state would be longer than would be inferred solely from steady-state considerations.
This transient behaviour has direct implications for tritium system design, as it informs both the time required to reach steady-state tritium throughput and the timescale over which the blanket inventory builds up.
In particular, the predicted response time provides guidance on the doubling time of the tritium inventory during reactor start-up and early operation~\cite{meschini_modeling_2023}.

To verify the FESTIM predictions via code-to-code comparison, the same transport problem was solved independently in OpenFOAM using the passive scalar formulation described in Section~\ref{subsec:HyT-meth}. 
Since the two codes solve the same governing equations, they are expected to yield comparable results. 
The temporal evolution of the outlet flux and blanket inventory from both codes is shown in figure~\ref{fig:plot_transient}.
The two solutions agree closely: The steady-state outlet flux is \SI{1.0}{mg.s^{-1}} in FESTIM and \SI{1.01}{mg.s^{-1}} in OpenFOAM, with corresponding inventories of \SI{243}{mg} and \SI{240}{mg}. 
The outlet flux reaches \SI{99.5}{\%} of its steady-state value after \SI{25.0}{min} in FESTIM and \SI{23.1}{min} in OpenFOAM; the small difference is most likely attributable to the different time discretisation schemes used by the two codes.
This close agreement between an independent finite volume solver and the finite element FESTIM model lends confidence to the predicted inventory and transient behaviour.

At steady state, the FESTIM outlet flux remains approximately \SI{1.5}{\%} below the neutronic production rate. 
By contrast, the finite volume OpenFOAM formulation conserves mass almost exactly, with a steady-state imbalance of only \SI{-0.00075}{\%}, reflecting its inherently conservative construction. 
As global conservation requires that outlet flux and production balance once a steady state is reached, the FESTIM residual is attributed to the stabilisation scheme: the artificial diffusion term, although necessary to suppress oscillations in the continuous Galerkin formulation at high Péclet numbers, is not strictly conservative. 
A promising route to remove this discrepancy is to adopt a discontinuous Galerkin discretisation, which is locally conservative by construction and stabilises advection-dominated transport through cell-wise upwind numerical fluxes. 
Such a scheme would eliminate the need for the artificial diffusion term and its tuning parameter, $\delta$, altogether and is identified as a priority for the workflow's future development.

From an extraction perspective, the tritium concentration in the outlet fluid is the key parameter, as the effectiveness of many extraction technologies depends directly on the local concentration, amongst other factors~\cite{Fuerst17112023}.
While the outlet flux characterises the overall tritium throughput, the outlet concentration governs the extraction system's efficiency and operating conditions.
The transient results indicate that the outlet concentration approaches its steady-state value on a timescale comparable to that of the outlet flux.
This highlights the importance of considering both outlet flux and outlet concentration when evaluating tritium transport and recovery in breeding blanket systems, particularly during reactor start-up or changes in operating conditions.

These considerations become particularly important for reactor concepts operating in pulsed or partially steady-state regimes.
Under such conditions, both the magnitude and temporal evolution of tritium concentrations and fluxes influence the performance of the extraction system.
Spatially and temporally resolved transport simulations, therefore, provide an essential tool for assessing tritium recovery strategies under realistic operating scenarios, beyond what can be inferred from steady-state metrics alone.

\begin{figure*}[t]
    \centering
    \includegraphics[width=0.8\linewidth]{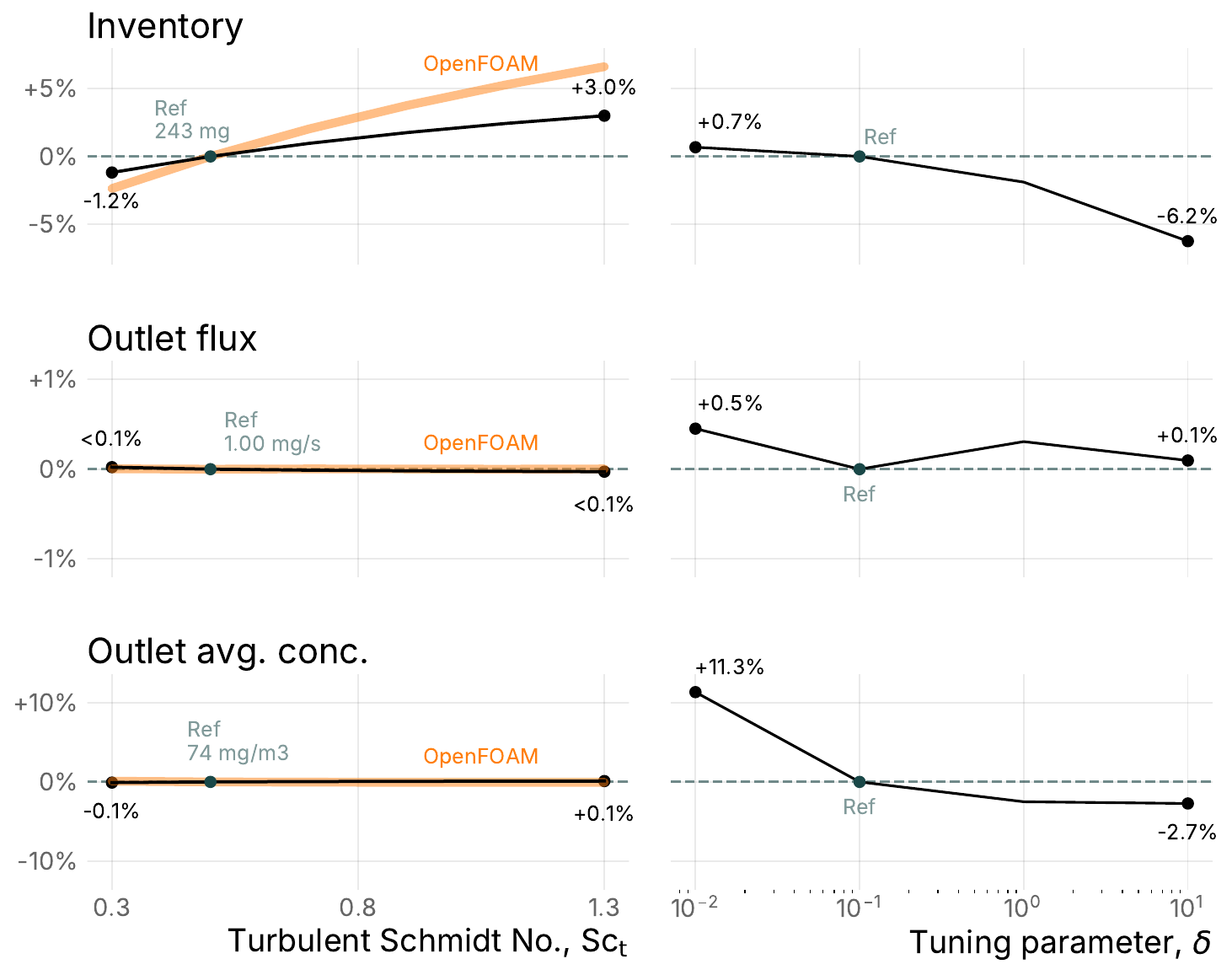}
    \caption{Influence of the numerical tuning parameter, $\delta$ (right), and the turbulent Schmidt number, $\mathrm{Sc}_t$ (left), on the steady-state blanket tritium inventory (top), outlet tritium flux (middle), and outlet-averaged concentration (bottom). Comparisons with results from OpenFOAM passive scalar transport (orange) for the turbulent Schmidt number are also shown.}
    \label{fig:plot_parametric}
\end{figure*}

To assess the robustness of the predicted tritium inventories and outlet behaviour, a limited parametric analysis was performed to examine the sensitivity of the results to key numerical and turbulence-related modelling parameters.
In particular, variations were considered in both the turbulent Schmidt number, Sc$_\mathrm{t}$, and the numerical tuning parameter, $\delta$.
The tuning parameter, $\delta$, controls the magnitude of the artificial diffusion term introduced to ensure numerical stability when solving advection-dominated transport problems at high Péclet numbers.
As described in Section~\ref{subsec:HyT-meth}, this tuning term introduces additional diffusion aligned with the local flow direction, analogous to SUPG or artificial diffusion methods, and is used in combination with the turbulent diffusivity to define the effective diffusion coefficient.
While necessary for numerical convergence, the tuning term can introduce artificial diffusion and numerical bias.
To assess its impact, $\delta$ was varied over a wide range, from 0.01 to 10, around a reference value of $\delta = 0.1$.
Particular attention was also given to the turbulent Schmidt number, Sc$_\mathrm{t}$, for which very limited data are available for liquid FLiBe flows.
Existing studies of turbulent scalar transport report a wide range of Sc$_\mathrm{t}$ values, typically between approximately 0.3 and 1.3, depending on flow conditions and fluid properties \cite{reynolds_prediction_1975, tominaga_turbulent_2007, yamamoto_direct_2015}.
For comparison, values of Sc$_\mathrm{t}$ around 0.7 are commonly adopted for water flows \cite{reynolds_prediction_1975}.
In the absence of a well-established value for FLiBe, a range of Sc$_\mathrm{t}$ values from 0.3 to 1.3 was therefore investigated to bound the expected physical uncertainty.

The influence of the numerical tuning parameter, $\delta$, is considered first. Reducing $\delta$ reduces artificial diffusion and brings the solution closer to the physically correct behaviour; across the investigated range, it leads to increased tritium retention, together with higher outlet fluxes and concentrations (figure~\ref{fig:plot_parametric}). 
Over this range, however, the resulting changes in inventory, outlet flux, and outlet concentration are modest, indicating that the predictions are relatively robust to this parameter. 
Reducing $\delta$ towards zero is nonetheless desirable for accuracy but was observed to introduce the first numerical instabilities near the interconnect and in flow stagnation zones, where concentration gradients are highest. 
This indicates that local mesh refinement in these regions is preferable to relying on artificial diffusion for stability.

The turbulent Schmidt number, $\mathrm{Sc}_t$, instead represents a physical uncertainty, as no consensus value exists for liquid FLiBe. 
Varying $\mathrm{Sc}_t$ across the range 0.3 to 1.3 again produces only modest changes in inventory, outlet flux, and outlet concentration (figure~\ref{fig:plot_parametric}).
The outlet flux in particular is only weakly sensitive to $\mathrm{Sc}_t$, indicating that transport at the outlet is dominated by advection, consistent with the high flow velocities in the breeder (figure~\ref{fig:res_foam_u}). 
Turbulent diffusion, therefore, mainly redistributes tritium within the breeder rather than governing its bulk removal.

Despite these insights, several limitations of the present model must be acknowledged.
Most notably, the simulations neglect tritium diffusion into and trapping within solid structural materials, which is expected to contribute to the overall tritium inventory.
Previous studies on the influence of trapping in structural materials (Eurofer) in a Water Cooled Lead Lithium blanket found a tritium inventory increase of approximately \SI{20}{\%} \cite{dark_influence_2021}.
However, this work neglected the potential impact of neutron damage on tritium retention.
This effect was found to increase model inventories in tungsten monoblocks by several orders of magnitude~\cite{hodille_tritium_2026, dark_modelling_2024}.
Therefore, a dedicated study focusing on trapping in structural materials will be required in the future.

In addition, permeation through the inner wall is not explicitly modelled, potentially allowing tritium to migrate from the breeder into plasma-facing components.
These effects are independent of the numerical and turbulence-related sensitivities discussed above and may contribute to tritium retention and redistribution.
Addressing these mechanisms will be an important focus of future work.

The present results represent an initial step towards fully resolved tritium transport modelling in ARC-class breeding blankets.
The analysis could be extended by incorporating a more detailed geometric description, including explicit treatment of solid walls and the associated tritium diffusion and trapping within structural materials.
Introducing a nonzero inlet concentration would enable closer coupling with realistic tritium extraction systems and allow direct assessment of extraction performance under transient operating conditions.
In addition, accounting for tritium implantation from the plasma into the first wall would provide a more complete representation of tritium sources and sinks within the blanket system.
Beyond purely transport-driven effects, future extensions could include chemical reactions within the liquid breeder, allowing investigation of tritium speciation and the influence of redox chemistry on transport and retention.

\section{Conclusion}

This work demonstrates a fully open-source, modular workflow for modelling tritium transport in an ARC-class molten salt breeding blanket at the component scale. 
Neutron transport, thermal hydraulics, and hydrogen isotope transport are coupled using OpenMC, OpenFOAM, and FESTIM, respectively, with data transfer enabled by the \texttt{openmc2dolfinx} and \texttt{foam2dolfinx} packages. 
This approach preserves solver specialisation while enabling transparent and reproducible multiphysics simulations.

Application of the workflow to a liquid immersion blanket geometry shows that tritium transport in the molten salt is dominated by flow-driven processes.
Turbulence-enhanced diffusion exceeds molecular diffusion by several orders of magnitude, leading to strong sensitivity of local tritium concentrations to the underlying flow structure. 
Regions of flow stagnation are associated with elevated tritium concentrations, while highly turbulent regions exhibit reduced accumulation due to enhanced mixing.
These results demonstrate the value of spatially resolved transport modelling for identifying unfavourable flow features and informing the optimisation of blanket design.

At the system level, the predicted outlet flux and the timescale to reach steady-state tritium throughput (approximately \SI{30}{min}) are broadly in line with previous ARC tritium cycle studies, while the predicted inventory is lower, reflecting differences in geometry and modelling assumptions. Furthermore, code-to-code comparisons with an OpenFOAM-based tritium transport solver within the FERMI workflow show excellent agreement and build confidence in the results.
The transient results further highlight the importance of outlet tritium concentration, in addition to mass flux, when assessing tritium extraction performance, particularly for pulsed reactor operation.

Sensitivity studies indicate that the predicted inventory and outlet behaviour are relatively robust to both the numerical stabilisation scheme and the turbulent Schmidt number, with the stabilisation parameter having the somewhat larger effect. 
The artificial diffusion introduced for numerical stability reduces tritium retention and leaves a small residual imbalance between the outlet flux and the neutronic production rate, indicating that local mesh refinement is preferable in regions of strong gradients. 
A key priority for future development is therefore to replace the stabilised continuous Galerkin formulation with a locally conservative discontinuous Galerkin scheme, which would stabilise advection-dominated transport through cell-wise upwind fluxes while removing the artificial diffusion term, its tuning parameter, and the associated conservation error. 
The turbulent Schmidt number remains a source of physical uncertainty, though its effect on the predicted outlet behaviour is modest, and improved characterisation of turbulence-driven scalar transport in molten salts would further constrain it.

The present model neglects tritium transport in solid structures, permeation through the inner wall, and chemical effects within the breeder, and assumes a perfectly efficient extraction system through the zero concentration imposed at the blanket inlet.
The predicted inventory should therefore be regarded as a lower bound: trapping in structural materials alone has been shown to increase inventories by approximately \SI{20}{\%} in comparable blanket systems~\cite{dark_influence_2021}, and neutron-damage-induced trapping may raise this considerably further~\cite{hodille_tritium_2026, dark_modelling_2024}.
Quantifying these contributions will require a dedicated study of trapping in irradiated structural materials.
Nevertheless, this study establishes a flexible and extensible framework for high-fidelity tritium transport modelling in ARC-class blankets, providing a foundation for future work incorporating solid domains, chemistry, different Li enrichment levels, a neutron multiplier, additional flow physics such as buoyancy and magnetohydrodynamics, and more realistic operating scenarios \cite{bonifetto_conceptual_2021}.

\section*{Acknowledgments}
This material is based upon work supported by the National Science Foundation under Grant No.~2449339.
Some of the authors of this paper are funded under Contract DE-FOA0002924 with the US Department of Energy.

The authors gratefully acknowledge Joseph Dean (University of Cambridge) for valuable discussions throughout this work, particularly on the discontinuous Galerkin formulation of the transport problem, which will be the focus of a forthcoming study.

\bibliographystyle{elsarticle-num}

\bibliography{references}

\end{document}